# Mikhail Vasil'evich Lomonosov

## *Panegyric to the Sovereign Emperor Peter the Great* [1]

***Panegyric to the Sovereign Emperor Peter the Great of blessed and everlastingly honored memory, prelected at the Solemn celebration of the coronation of Her Imperial Majesty, the Supremely August, Most Autocratic, Great Sovereign Empress Elizaveta Petrovna, Autocrat of All Russia***

[ *Panégyrique de Pierre le Grand, prononcé dans la séance publique de l'Académie impériale des sciences le 26 avril 1755 Par Mr. Lomonosow* ]

[ *Слово похвальное блаженные пямяти государю Императору Петру Великому, говоренное апреля 26 дня 1755 года* ]

My listeners, as we commemorate the most holy anointing and coronation of our most gracious Autocrat as Sovereign of the All-Russian state[2], we behold God's favor to Her and to our common fatherland, favor like unto that at which we marvel in Her birth and in Her coming into possession of Her patrimony. Her birth was made wonderful by omens of empire; Her ascension to the throne is made glorious by courage protected from on high; Her acceptance from the Lord's hand of Her Father's crown and splendid victories is infused with reverential joy. It may be that there are still those who are in doubt whether Rulers are set up on earth by God or whether they obtain their realms through chance; yet in itself the birth of our Great Sovereign compels conviction here, inasmuch as even at that time She had been elected to rule over us. Not the dubious divinations of astrology, based on the position of the planets, nor yet other manifestations or changes in the course of nature, but clear signs of God's Providence shall serve as proof of this. PETER'S most glorious victory over His foes at Poltava took place in the same year as the birth of this Great Daughter of His, and the Conqueror riding in triumph into Moscow was met by the coming into the world of ELIZABETH[3]. Is this not the finger of Providence? Do we not hear a prophetic voice in the mind's ear? Behold, behold the fruition of that bliss promised to us in auguries. PETER triumphed in victory over external enemies and in the extirpation of traitors[4]; ELIZABETH was born for similar triumphs. PETER, having returned the crown to the lawful Sovereign[5], made His entry into His Father's city; ELIZABETH entered the human community in order that Her father's crown should afterward revert to Her. PETER, having protected Russia from pillage, brought joy, secure and serene, in place of gloomy fear; ELIZABETH saw the light of day that She might pour the sunshine of happiness over us and free us from the gloom of sorrows. PETER led in His train numerous captives, conquered no less by His magnanimity than by His courage[6]; ELIZABETH was delivered from the womb that She might thereafter captivate the hearts of Her subjects with philanthropy, gentleness, and generosity. What wondrous destinies ordained by God do we behold, O ye who hearken to me! Victory together with birth, the delivery of the fatherland with the delivery of the child from the Mother, the extraordinary triumphal procession and the ordinary rites of birth, swaddling clothes together with triumphal laurels, a baby's first cry mingled with joyous applause and acclamation! Were not all these things an earnest of Her Father's virtues and of Her Father's realm to the child ELIZABETH who was then born?

---

[1] Updated English translation of Ref.[1]; footnotes and commentary by V.Shiltsev
[2] Empress Elizaveta Petrovna, the Empress of Russia from 1741 until her death in 1761, daughter of Peter the Great and Catherine I, was crowned on April 25 (o.s., May 6, n.s.,) 1742.
[3] Empress Elizaveta Petrovna was born on December 18(29), 1709, on the day originally planned by Peter the Great for triumphant entry to Moscow to celebrate his victory over Swedish Charles XII in the Battle of Poltava (July 27). Family celebration took precedence and the triumph was moved by three days to December 21.

   How much Almighty Providence did lend aid to Her own heroism on the path to the throne is a matter whereof joyous memories shall not grow silent throughout the ages. For it was moved by the power and spirit of Providence that our Heroine brought salvation and renewal to the All-Russian state, to its observed glory, to the great deeds and designs of Peter, to the inner satisfaction of our hearts, and to the general happiness of a large part of the world. To rescue a single man is a great deed; but how immeasurably greater is the salvation of an entire people! In thee, our dear fatherland, in thee do we behold abundant examples of this. Angered by the internecine feuds, perfidies, pillaging, and fratricides of our forefathers, God once bound thee in slavery to a foreign people, and on thy body, stricken with deep wounds, He laid heavy fetters[7]. Then, moved by thy groans and lamentations, He sent thee valorous Sovereigns, liberators from slavery and suffering, who, joining together thy shattered limbs, did restore and increase thy former power, majesty, and glory. From no lesser downfall has the Russian people been rescued by the Great ELIZABETH, whom God has brought to Her Father's throne; but in a fashion even more deserving of astonishment. Internal diseases could be more calamitous than external; likewise a danger nurtured in the bosom of a state is more harmful than attacks from without[8]. External sores are healed more easily than internal injuries. But when we compare the healing of Russia from defeat inflicted externally by the weapons of barbarians with the astonishing cure of a hidden internal ailment accomplished by ELIZABETH'S hand, we find the reverse. In the past, in order to heal external wounds, fields and rivers were empurpled no less with Russian than with Agarian blood[9]. In our blessed days great-hearted ELIZABETH swiftly, without any suffering on our part, destroyed the disease which had taken root inside Russia, and it was as if She did heal the ailing fatherland with a single word filled with divine power, saying: *Arise and walk, arise and walk, Russia*[10]. *Shake of thy doubts and fears; and filled with joy and hope, stand forth, exult, and rise to eminence.*

---

[4] Reference to Ivan Mazepa (1639-1709), Hetman (military commander) of Zaporozhian Cossacks in 1687-1708, who quisled Peter and sided with King Charles XII of Sweden in 1708, got anathemed by the Russian Orthodox church and died shortly after the Poltava battle.

[5] As a result of the Poltava victory, Peter's ally the Saxony Elector Augustus II, who had previously been removed from the Polish throne by Charles XII, was restituted.

[6] Peter the Great was remarkably kind to 19,000 Swedes, captured as the result of the Poltava battle, and even called out the top captured Swedish military commanders to the celebratory feast after the battle.

[7] Reference to two and half century Mongolian rule over Russian principalities over 13th to 15th centuries.

[8] Reference to tough 1730-1740 government by a Courland Johann von Biron (1690-1772), a favorite of the Russian Empress Anna Ioannovna, that was considered by many as pro-German and anti-Russian.

Such are the images evoked in our thoughts, O Listeners, by memory of the joy then aroused! But they are magnified when we consider that we were then freed not only from oppression but also from contempt. How did the nations judge us before our delivery? Do not their words still echo in our memory? The Russians, the Russians have forgotten PETER the Great! They do not repay His labors and services with due gratitude; they do not install His Daughter on Her Father's throne; She is abandoned and they do not help Her; She is rejected and they do not restore Her; She is scorned and they do not avenge Her. Oh, how great is the shame and derision! But the incomparable Heroine has by Her ascension to the throne[11] removed disgrace from the sons of Russia and shown to the whole world the truth—that it was not that our ardor was lacking, but that Her magnanimity was all-enduring; that it was not that our zeal failed, but that She did not wish the spilling of blood; that it was not to our faintheartedness that these things must be ascribed but to God's Providence, which deigned thereby to show His power and Her courage and to redouble our joy. Such benefactions did the Almighty ordain for us by the ascension of Great ELIZABETH to Her Father's throne! Then what is our present celebration? The summit and crown of the things heretofore mentioned. The Lord has crowned Her wondrous birth, has crowned Her glorious ascension, has crowned Her incomparable virtues. He has crowned Her with grace, heartened Her with hopeful joy, and blessed Her with resounding victories, victories which resemble Her ascension. For just as internal enemies have been conquered without bloodshed[12], so also have external enemies been overcome with little loss[13].

Our Monarch is robed in purple, is anointed for imperial rule, is crowned and receives the Scepter and the Orb. Russians rejoice, filling the air with applause and acclamation. Enemies quail and blench. They slip away, turning their back on the Russian army, hide beyond rivers, mountains, swamps; but everywhere the strong hand of crowned ELIZABETH hems them in; from Her magnanimity alone do they receive relief. How clear are the auguries of Her blessed rule, which we see in all that has been said above, and how we joyously marvel at their longed-for realization! Following the example of Her great progenitor, she gives crowns to Sovereigns, calms Europe with peaceful arms[14], consolidates the Russian inheritance[15]. Gold and silver flow out of the bowels of the earth[16], giving pleasure to Her and to the community; subjects are relieved of their burdens; neither within the realm nor beyond its bounds is the soil stained with Russian blood; the people multiply and revenues increase; magnificent buildings arise; the courts are reformed; the seeds of learning are planted in the state, while everywhere there reign dearly loved peace and quiet and an age which resembles our Monarch Herself.

---

[9] reference to victory over Mongols in the Kulikovo field battle (1380) which the beginning of the end of the Golden Horde rule. Agarians, Hagarenes, the descendants of Hagar - a biblical person in the Book of Genesis who was an Egyptian slave of Sarah and were given to Abraham to bear a child - a term widely used by early Syriac, Greek, Coptic and Armenian sources to describe the early Arab conquerors of Mesopotamia, Syria and Egypt and came also to mean any Muslim.

[10] Reference to Mark 2:9

[11] Elizabeth came to power as a result of a palace coup, having removed Ivan V Antonovich from the throne.

[12] A hint to replacement of punishment of the previously powerful H.J.Ostermann, B.C. Munnich, M.G. Golovkin, K.G. Lowenwolde from execution to Siberian exile in January 1742

[13] Reference to the successful end of the 1741-1743 war with Sweden.

And so, since our incomparable Sovereign has raised up Her Father's realm as presaged by birth, gained by courage, confirmed by Her triumphant coronation, and adorned by glorious deeds, She is in justice the true Heiress of all His deeds and praise. Therefore in praising PETER we shall praise ELIZABETH.

Long ere this the arts and sciences should have represented His glory in clear portrayals; long have they wished to laud the incomparable deeds of their Founder[17] at a special grand assembly, but, knowing how great an art is required to frame a speech worthy of the subject, they have remained silent until this day. For of this Hero must be said what has never yet been of others. There is none equal to Him in deeds, nor is there any precedent in rhetoric which the thought might follow to plumb without mishap the depths of the vast number and grandeur of His deeds. But it was finally decided that it was better to show a deficiency of eloquence than of gratitude, that it was better to compose a discourse graced by sincerity and couple it with words of converse spoken in studied simplicity than to remain in silent amid so much acclamation on this festive occasion. And this is especially so since the Almighty Lord has enhanced the beauty of all our celebrations by sending in the person of the young Sovereign, the Grand Duke PAVEL PETROVICH[18], a universally longed-for pledge of His divine favor to us, which we see in the continuation of PETER'S line. And so, abandoning timorous doubt and yielding to zealous daring, we must use, or rather exhaust, whatever spirit and voice we have in praise of our Hero. As I embark on this undertaking, with what shall I begin my discourse? With His bodily endowments? With the greatness of His strength? But it is manifest in His mastery of burdensome labors, labors without number, and in the overcoming of terrible obstacles. Shall I begin with His heroic appearance and statute combined with majestic beauty? But apart from the many who vividly call to mind an image of Him engraved in their memory, there is the witness those in various states and cities who, drawn by His fame, flocked out to admire a figure appropriate to His deeds and befitting a great Monarch. Should I commence with His buoyancy of spirit? But that is proved by the tireless vigilance without which it would have been impossible to carry out deeds so numerous and great. Wherefore I do immediately proceed to present these deeds, knowing that it is easier to make a beginning than to reach the end and that this Great Man cannot be better praised than by him who shall enumerate His labors in faithful detail, were it but possible to enumerate them.

---

[14] Movement of the Russian expeditionary force to the Rhine accelerated the 1748 Aachen peace treaty to end the five-year war for the Austrian inheritance.

[15] Peter's grandson, grandson of Charles XII sister and Elizabeth's nephew Petr Fedorovich (future Emperor Peter III) was announced heir to the throne in February 1742 and that was generally perceived as a guarantee of stability of power.

[16] the first native gold and silver deposits were discovered in the Urals in 1745-1746. Since the late 1740s gold and silver mining began in Altai mountains and Trans-Baikalia.

[17] Peter the Great was the founder of the Academy of Sciences in 1724. Updated Academy's Charter (1747) prescribed three public assemblies of the Academy every year and the first one in January should be in the memory of the founder. In reality that did not happen until April 1755. After delays, the first public assembly in the Academy was set only for September 6, 1749, on the Elizabeth's namesday. At that time Lomonosov was instructed to prepare and read his "Panegyric to the Empress Elizabeth".

[18] Grand Prince Pavel Petrovich (1754-1801) – future Emperor Pavel I, son of Petr Fedorovich and his wife Ekaterina Alekseevna (future Empress Catherine II) - was born on September 20, 1754.

And so, to the extent that strength and the brevity of limited time will permit, we shall mention only His most important deeds, then the mighty obstacles therein overcome, and finally the virtues which aided Him in such enterprises.

As a part of His grand designs the all-wise Monarch provided as a matter of absolute necessity for the dissemination of all kinds of knowledge in the homeland, and also for an increase in the numbers of persons skilled in the higher branches of learning, together with artists and craftsmen; though I have given His paternal solicitude in this matter the most prominent place, my whole speech would not be long enough to describe it in detail. For, having repeatedly made the rounds of the European states like some swift soaring eagle, He did induce (partly by command and partly by His own weighty example) a great multitude of His subjects to leave their country for a time and to convince themselves by experience how great all an advantage a person and an entire state can derive from a journey of inquiry in foreign regions. Then were the wide gates of great Russia opened up; then over the frontiers and through the harbors, like the tides in the spacious ocean, there did flow in constant motion, in the one direction, the sons Russia, journeying forth to acquire knowledge in the various sciences and arts[19], and, in the other direction, foreigners arriving with various skills, books, and instruments. Then to the study of Mathematics and Physics, previously thought of as forms of sorcery and witchcraft, but now arrayed in purple, crowned with laurels, and placed on the Monarch's throne, reverential respect was accorded in the sanctified Person of PETER. What benefit was brought to us by all the different sciences and arts, bathed in such a glow of grandeur, is proved by the superabundant richness of our most varied pleasures, of which our forefathers, before the days of Russia's Great Enlightener, were not only deprived but in many cases had not even any conception. How many essential things which previously came to Russia from distant lands with difficulty and at great cost are now produced inside the state, and not only provide for our needs but also with their surplus supply other lands. There was a time when the neighbors on our borders boasted that Russia, a great and powerful state, was unable properly to carry out military operations or trade without their assistance, since its mineral resources included neither precious metals for the stamping of coins nor even iron, so needful for the making of weapons with which to stand against an enemy. This reproach disappeared through the enlightenment brought by PETER; the bowels of the mountains have been opened up by His mighty and industrious hand. Metals pour out of them, and are not only freely distributed within the homeland but are also given back to foreign peoples as if in repayment of loans. The brave Russian army turns against the enemy weapons produced from Russian mines by Russian hands[20].

---

[19] Over thirty years of absolute reign (1796-1725), Peter had sent abroad a thousand of (mostly young) Russians for studies in Holland, France, England, Italy, etc.

In the establishment of the sizable army needed for the defense of the homeland, the security of His subjects, and the unhindered carrying out of important enterprises within the country, how great was the solicitude of the Great Monarch, how impetuous His zeal, how assiduous His search of ways and means! Since all these things exceed our capacity for admiration, shall we be able to describe them in words? The sire of our all-wise Hero, the Great Sovereign Tsar ALEKSEI MIKHAILOVICH[21] of blessed memory, among other glorious deeds, laid the foundation of a regular army, with which He had great success in war, as witnessed by His victorious campaigns in Poland and by the provinces restored to Russia. But all His attentiveness to military matters was cut short by His death. The old disorders returned, and the Russian army was better able to show its strength in numbers than in skill. How much this strength later decreased is clear from the unavailing military undertakings of the time against the Turks and Tatars[22] and, most of all, from the unbridled and ruinous mutinies of the strel'tsy[23] which were the result of lack of proper discipline and organization. In such circumstances who could dream that a twelve-year-old youth, kept away from the conduct of state affairs and shielded from evil only by the sagacious protection of His affectionate Mother, amid ceaseless terrors, and amid spears and lances drawn against His relatives and well-wishers and against Himself, could have begun to establish a new regular army, whose might the enemies were to feel very soon afterward. They felt it and trembled, and now the entire universe is justly amazed by it. Who could imagine that from what seemed to be a child's game so great an enterprise could arise? Some people, seeing a few young men in company with their young Sovereign handling, light weapons in various ways, were of the opinion that this was nothing more than an amusement to Him, wherefore these new recruits were called Toy Soldiers. Others, being more perceptive and noticing the bloom of heroic high spirits on the Youth's face, the intelligence shining from his eyes, and the quick authority of His movements, reflected that Russia could then already expect a valiant Hero, a Great Monarch. But to recruit many great regiments of infantry and cavalry, to provide all with clothing, pay, weapons, and other military stores, to teach them [His] new Articles of War, to establish properly constituted field and siege artillery (for which no little knowledge of Geometry, Mechanics, and Chemistry is needed), and above all to have skilled commanders everywhere—all this might justly have seemed an impossibility; for a notable lack of all these means and loss of Sovereign power had removed the least hope and remotest probability of such a thing. Yet what was it that ensued? Contrary to general popular expectation, contrary to the disbelief of those who had abandoned hope, and in the teeth of the hostile intrigues and venomous murmurings of Envy itself, suddenly PETER'S new regiments thundered forth, arousing joyous hope in loyal Russians, fear in their opponents, and amazement on both sides. The impossible was made possible by extraordinary zeal, and above all by an unheard-of example. In former times the Roman Senate, beholding the Emperor Trajan standing before the Consul to receive from the dignity of Consul, exclaimed: "Through this thou art the greater, the more majestic!"[24] What exclamations, what applause were due to PETER the Great for His unparalleled self-abasement? Our fathers beheld their crowned Sovereign not among the candidates for a Roman consulship but in the ranks of common soldiers, not demanding power over Rome, but obedient to the bidding of His subjects. O you beautiful regions, fortunate regions which beheld a spectacle so wondrous! Oh, how you marveled at the friendly contest of the regiments of a single Sovereign, both commander and subordinate, giving orders and obeying them! Oh, how you admired the siege, defense, and capture of new Russian fortresses, not for immediate mercenary gain but for the sake of future glory, not for putting down enemies but to encourage fellow countrymen. Looking back at those past years, we can now imagine the great love for the Sovereign and the ardent devotion with which the newly instituted army was fired, seeing Him in their company at the same table, eating the same food, seeing His face covered with dust and sweat, seeing that He was no different from them, except that in training and in diligence He was superior to all. By such an extraordinary example the most wise Sovereign, rising in rank alongside His subjects, proved that Monarchs can in no other way increase their majesty, glory, and eminence so well as by such gracious condescension. The Russian army was toughened by such encouragement, and during the twenty years' war with the Swedish Crown[25], and later in other campaigns, filled the ends of the universe with the thunder of its weapons and with the noise of its triumphs. It is true that the first battle of Narva was not successful[26]; but the superiority of our foes and the retreat of the Russian army have, through envy and pride, been exaggerated to their glorification and our humiliation, out of all proportion to the actual event. For although most of the Russian army had seen only two years' service and faced a veteran army accustomed to battle, although disagreement arouse between our commanders, and a malicious turncoat[27] revealed to the enemy the entire position in our camp, and Charles XII by a sudden attack did not give the Russians time to form ranks—yet even in their retreat they destroyed the enemy's willingness to fight on to final victory. Thus the only reason the Russian Life Guard, which had remained intact, together with another sizable part of the army, did not dare to attack the enemy thereafter was the absence of its main leaders, who had been summoned by Charles for peace talks and detained as prisoners[28]. For this reason the Guards and the rest of the army returned to Russia with their arms and war chest, drums beating and banners flying. That this failure occurred more through the unhappy circumstances described than through any lack of skill in the Russian troops and that PETER'S new army could, even in its infancy, defeat the seasoned regiments of the enemies, was proved in the next year[29] and subsequently by many glorious victories won over them.

---

[20] Under Peter's industrial policies, the number of metallurgical plants more quintupled to 88, the export of iron began in 1716 and over the 20 years increased hundred times to 3200 tons, 9 firearms and cold-arms factories were set up.

[21] Peter's father, Tzar Aleksei Mikhailovich (1629-1676), reigned from 1645.
[22] Reference to unsuccessful Crimean campaigns of Prince. V.V.Golitsyn in 1686 and 1688
[23] The units of Russian elite firearm infantry, established by Ivan the Terrible in mid-16th century. Their 1698 mutiny was mercilessly crushed by Peter, then they were technically abandoned and fully ceased by 1720s.

---

[24] Amended quote from Pliny Jr.'s panegyric to Trajan on the third election as a consul: "Ingens, Caesar, et par gloria tua".

[25] The Northern War with Sweden (1700-1721).
[26] On November 19, 1700 the Russian army suffered a heavy defeat in the battle of Narva.
[27] Turncoat Estland-born captain of the Preobrazhensky regiment Jacob Gummert ran to the Swedes.

[28] Seven Russian generals and seven colonels were captured and detained.

It is to you, our now peaceful neighbors[30], that I address my words: When you hear these praises of our Hero's deeds of war and my eulogies of the Russian army's victories over you, consider yourselves honored rather than disgraced. For to stand long against the mighty Russian people, to stand against PETER the Great, against the Man sent by God to astound the universe, and in the end to be conquered by Him is more glorious than to conquer weak armies under poor leadership. Justly consider the boldness of your hero Charles to be your true glory; and assert, with the whole world's agreement, that there was scarcely any man who could hold out in the face of his wrath, had not God's wondrous Providence set up PETER the Great against him in our fatherland. PETER'S bold regiments, organized on the basis of a regular army, proved by a quick succession of victories what ardent zeal and what great skill in the art of war they had acquired from wise instruction and example. Leaving aside the many victories, which the Russian army had learned to count by the number of its battles, not mentioning the great number of captured towns and stout fortresses, we have sufficient witness in the two main victories at Lesnaya and Poltava[31]. Where else did the Lord manifest greater favor to us? Where else was clearer revelation given of [how great were] the successes achieved by PETER'S blessed initiative and zealous ardor in instituting the new army? What more miraculous or improbable event could have ensued? An army with a long-standing regular tradition, which had been brought from enemy lands renowned for their battle prowess under the leadership of commanders who had spent all their time in the practice of war, an army lavishly equipped with all kinds of weapons, shunned battle with the new Russian regiments, greatly inferior in numbers. But the latter, giving their opponents no rest, by swift movement overtook them, fought, and were victorious; and the enemy's chief commander barely escaped captivity with a few remnants to take the dolorous news to his Sovereign. Although the latter was greatly disturbed, yet in the confidence of his bold and impetuous spirit he still spurred himself on against Russia. He still could not be convinced that PETER'S young army could stand against his matured forces advancing under his own leadership and, putting his trust in the audacious encouragement of a conscienceless traitor to Russia[32], did not hesitate to enter the border regions of our homeland. He tried to win over Russia with arrogant arguments and already considered that the whole North was beneath his heel. But God, in reward for unceasing labors, granted PETER a complete victory over this contemner of His efforts, who, contrary to his expectations, became eyewitness of our Hero's unbelievable successes in warfare and could not even in flight escape from the haunting thoughts of Russia's steady courage.

---

[29] In June 1701, two Swedish ships were captured near Arkhangelsk.

[30] i.e., Swedes.
[31] On October 9, 1708 and July 8, 1709, correspondingly.
[32] Mazepa, see footnote 4.

Having covered Himself and His army with glory throughout the world by such famous victories, the Great Monarch finally proved that he had been at pains to establish His army mainly in the interests of our safety! For He decreed that it should never be dispersed, even in times of untroubled peace (as had happened under previous Sovereigns, frequently to no little loss of the country's might and glory), and also that it should always be kept in proper readiness. Oh, truly paternal solicitude! Many times did He remind the loyal subjects close to Him, sometimes tearfully embracing and begging them, that the renewal of Russia—undertaken at such great pains and with such marvelous success—and most particularly the art of war, should not be neglected after Him. And at the very time of general rejoicing, when God had blessed Russia with a glorious and advantageous peace with the Swedish Crown, when ardent felicitations and the deserved titles of Emperor, "the Great," and "Father of His country" were being offered[33], He did not lose the opportunity to impress publicly on the Governing Senate that, though one may hope for peace, one must not grow slack in military matters. Did He not hereby give a clear sign that these high titles were not pleasing to Him if not accompanied by the preservation and maintenance of a regular army forever in the future?

Having cast a quick glance over PETER'S land forces, which came to maturity in their infancy and combined their training with victories, let us extend our gaze across the waters, my Listeners; let us observe what the Lord has done there, His marvels on the deep, as made manifest by PETER to the astonishment of the world.

The far-flung Russian state, like a whole world, is surrounded by rat seas on almost every side and sets them as its boundaries. On all of them we see Russian flags flying. Here the mouths of great rivers and new harbors scarcely provide space for the multitude of craft; elsewhere the waves groan beneath the weight of the Russian fleet, and the sounds of its gunfire echo in the chasms of the deep. Here gilded ships, blooming like spring, are mirrored on the quiet surface of the waters and take on double beauty; elsewhere the mariner, having reached a calm haven, unloads the riches of faraway countries to give us pleasure. Here new Columbuses hasten to unknown shores to add to the might and glory of Russia[34]; there a second Tiphys[35] dares to sail between the battling mountains[36]; she struggles with snow, with frost, with everlasting ice, desirous to unite East and West. How did the power and glory of the Russian fleets come to be spread over so many seas in a short time? Whence came the materials, whence the skill? Whence the machines and implements needed in so difficult and varied an enterprise? Did not the ancient giants tear great oaks from dense forests and lofty mountains and throw them down for building on the shores? Did not

---

[33] After the conclusion of the Nystad peace treaty with Swedish Empire favorable to Russia (1721), the Governing Senate solemnly requested Peter the Great to accept the titles of "Emperor of All Russia", "Peter the Great" and "Father of the Fatherland".
[34] Reference to Christopher Columbus (1451-1506) and Russian navigators who discovered land in the Far East during the first and second Kamchatka expeditions (1725-1730, 1733-1743).
[35] In ancient Greek mythology, the helmsman of the Argonauts.
[36] Icebergs.

Amphion[37] with sweet music on the lyre move the various parts for the construction of those wondrous fortresses which fly over the waves? To such fancies would PETER'S wondrous swiftness in building a fleet truly have been ascribed if an exploit so improbable and seemingly beyond human strength had been performed in far-off ancient times, and if it had not been fixed in the memory of many eyewitnesses and in unexceptionably reliable written records. In the latter we read with amazement, while from the former in friendly conversation we hear—and not without emotion—that it cannot be determined whether PETER the Great gave more pains to the founding of His army or His navy. However, there is no doubt that He was tireless in both and exemplary in both. For, seeking knowledge of everything which might happen in land battles, He not only went through all the ranks Himself, but also tried out for Himself all the crafts and kinds of work, that He might not overlook a neglect of duty on the part of anyone nor require of anyone efforts beyond his powers. In the navy, too, there was nothing left untouched, untried by His keen mind and industrious hands. From that very time when the contriving of a boat (which, though small in dimensions, was great in influence and fame) aroused in PETER'S unsleeping spirit the salutary urge to found a fleet and to show forth the might of Russia on the deep, He applied the forces of His great mind to every part of this important enterprise. As He investigated these parts, He became convinced that in a matter so difficult there was no possibility of success unless He Himself acquired adequate knowledge of it. But where was that to be obtained? What should the Great Sovereign undertake? It had formerly been a sight to behold when an immense throng of people poured out to see the entrancing spectacle on the fields of Moscow, as our Hero, still little more than a child—in the presence of the entire imperial household, of illustrious dignitaries of the Russian state, and of an illustrious gathering of the nobility, now joyful, now fearful of harm to His health—labored at laying out a regular fortress[38] like a craftsman, digging moats and piling up earth for embankments behind the ramparts like a common soldier, giving orders to all as a Sovereign, setting an example to all, like an all-wise Teacher and Enlightener. But greater still was the amazement that He aroused, greater the spectacle that He presented to the eyes of the whole world when, becoming convinced of the untold benefits of navigation—first on the small bodies of water in the Moscow area, then on the great breadth of Lake Rostov and Lake Kubinsk, and finally on the expanse of the White Sea[39]—He absented himself for a time from His dominions and, concealing the Majesty of His Person among humble workmen in a foreign land, did not disdain to learn the shipwright's craft. Those who chanced to be His fellow-apprentices at first marveled at the amazing fact that a Russian had not only mastered simple carpentering work so quickly, had not only brought Himself to the point where He could make with His own hands every single part needed in the building and equipping of ships, but had also acquired such skill in marine architecture that Holland could no longer satisfy His deep understanding. Then how great was the amazement that was aroused in all when they learned that this was no simple Russian, but the

---

[37] Son of Zeus and Antiope, famous for his wondrous play on the lyre and employing its magic to build the walls of Thebes.

[38] Presburg fortress, built by young Peter in 1684 in Preobrazhensky village for training purpose.

[39] Peter's first voyage along the White Sea took place in August 1693.

Ruler of that great state Himself who had taken up heavy labors in hands born and anointed to bear the Scepter and the Orb. But was it merely out of sheer curiosity or, at the most, for purposes of instruction and command, that He did in Holland and Britain attain perfection in the theory and practice of equipping a fleet and in navigational science? Everywhere the Great Sovereign aroused His subjects to labor, not only by command and reward, but also by His own example! I call you to witness, O great Russian rivers; I address myself to you, O happy shores, sanctified by PETER'S footsteps and watered by His sweat. How many times you resounded with high spirited and eager cries as the heavy timbers, ready for launching of the ship, were being slowly moved by the workmen, and then, at the touch of His hand, made a sudden spurt toward the swift current, inspiring the multitude, encouraged by His example, to finish off the huge hulks with incredible speed. To what a marvelous and rousing spectacle were the assembled people treated as these great structures moved nearer to launching! When their indefatigable Founder and Builder, now moving topside, now below, now circling round, tested the soundness of each part, the power of the machinery, and the precision of all the preparations and, by comma lid, encouragement, ingenuity, and the quick skill of His tireless hands, rectified the defects which He had detected. In this unflagging zeal, this invincible persistence in labor, the legendary prowess of the ancients was shown in PETER'S day to have been not fiction but the very truth!

What pleasure was afforded the Great Sovereign by the successes in maritime affairs achieved through His zeal, to the indescribable benefit and glory of the state, may easily be seen from the fact that He not only he stowed awards on those who had labored with Him but also gave insensitive timber a glorious token of His gratitude. The Neva's waters air covered with vessels and flags; its banks cannot hold the great multitude of the assembled spectators; the air vibrates and groans with the people's exclamations, with the noise of oars, the voice of trumpets, and the sound of igniferous monsters. What happiness, what joy does heaven semi us! Whom does our Monarch come out to greet in such magnificence? A decrepit little boat![40] But one that occupies pride of place in the new and mighty Fleet. Contemplating the majesty, beauty, might, and glorious deeds of the one and at the same time the smallness and feebleness of the other, we see that no one in the world could have achieved such things without PETER'S titanic daring in enterprise and tireless vigor of execution.

Exemplary on land, peerless on the waters in His might and military glory was our Great Defender!

From this brief account, which contains but a small part of His labors, I already feel fatigue, O my Listeners; yet I see the great and far-reaching field of His merits before me! And so that my strength and limited time may suffice to complete the drift of my discourse, I shall make all possible haste.

---

[40] Reference to Peter's Botik (also called St. Nicholas), a miniaturized scaled-down English warship discovered by young tzar at the Royal Izmaylovo Estate in 1688, restored by master Carsten Brandt, and used by Peter to learn to sail on the waters near Moscow. The symbolic celebration of the Botik as the "grandfather of the Russian fleet" took place on August 10, 1723 in Kronstadt. It is now enshrined in St. Petersburg.

For the foundation and bringing into action of so great a naval and land force, and also for the construction of new towns, fortresses, and harbors, for the joining of rivers with great canals, for the strengthening of frontiers with ramparts, for long-lasting war, for frequent and distant campaigns, for the construction of public and private buildings in a new architectural style, for the finding of experienced persons and all other means for the dissemination of science and the arts, for the maintenance of new court and state officials---how vast a treasury was needed for these things anyone can easily imagine, and conclude that the revenues of PETER'S Forefathers could not suffice for this. Wherefore the sagacious Sovereign did strive most earnestly to increase internal and external state revenues without ruining the people. And He had the native wit to perceive that by means of a single institution not only would great gains accrue to the treasury, but the general tranquility and safety of His subjects would also be assured. For at a time when the total number of the Russian people and the place of residence of each person were not yet known, there was no curb on arbitrary conduct, and no one was forbidden to change his place of residence or to wander about as the whim took him. The streets were filled with shameless, loafing beggars; the roads and great rivers were often blocked by the thieves and by whole regiments of murderous brigands who brought ruin to towns and villages alike. The wise Hero converted harm into benefit, laziness into industriousness, pillagers into defenders; when He had counted the multitude of His subjects, He bound each one to his dwelling and imposed a light, but fixed, tax; in this way the internal revenues of the treasury were increased, and a definite amount of such revenue was assured—and, likewise, the number of persons on recruiting lists. Industriousness and strict military training were also increased[41]. Many who, under previous conditions, would have been dangerous robbers He compelled to be ready to die for their country.

I say nothing of the assistance afforded in this matter by other wise institutions, but will mention the increase of external revenues. Divine Providence aided the good designs and efforts of PETER, through His hand opening new ports on the Variagian Sea[42] at towns conquered by His valor and erected by His own labors. Great rivers were joined for the more convenient passage of Russian merchants, duty regulations were established, and commercial treaties with various peoples were concluded. What benefit proceeded from the growth of this abundance within and without has been clear from the very foundation of these institutions, for while continuing to fight a burdensome war for twenty years Russia was free from debts[43].

---

[41] Reference to Petrine censuses of 1710, 1716-1717 and 1718-1724 to strengthen fiscal practices. The first one accounted 20% decrease of the number of taxed peasant households compared to the 1678 census. To reverse the trend which continued in 1716-1717, several measures were ordered to keep the people at their land, avoid unaccounted and untaxed wanderers and assure (rather heavy) tax revenues. Per capita instead of the household tax was introduced in 1717 and initially amounted to 80 kopecks, by the end of the reign - 74 kopecks.

[42] Baltic Sea; old Slavonic Variagi is for Varangians, Vikings.

What, then, have all PETER'S great deeds already been depicted in my feeble sketch? Oh, how much labor still remains for my thoughts, voice, and tongue! I ask you, my Listeners, out of your knowledge to consider how much assiduous effort was required for the foundation and establishment of a judiciary, and for the institution of the Governing Senate, the Most Holy Synod, the state collegiums[44], the chancelleries, and the other governmental offices with their laws, regulations, and statutes; for the establishment of the table of ranks[45] and the introduction of decorations as outward tokens of merit and favor; and, finally, for foreign policy, missions, and alliances with foreign powers. You may contemplate all these things yourselves with minds enlightened by PETER. It remains to me only to offer a brief sketch of it all. Let us suppose that before the beginning of PETER'S enterprises someone had happened to leave his native Russia for distant lands where His name had not thundered forth—if such a land there be on this earth. Returning later to Russia, he would see new knowledge and arts among the people, new dress and customs, new architecture and household furnishings, newly built fortresses, a new fleet, and a new army; he would see not only the different aspect of all these things but also a change in the courses Of rivers and in the boundaries of the seas. What would he then think? He could come to no other conclusion than that he had been on his travels for many centuries, or that all this had been achieved in so short a time by the common efforts of the whole human race or by the creative hand of the Almighty, or, finally, that it was all a vision seen in a dream.

From these words of mine, which reveal scarcely more than the mere shadow of PETER'S glorious deeds, it may be seen how great they are! But what is one to say of the terrible and dangerous obstacles encountered on the path of His mighty course? They have exalted His honor the more! The human condition is subject to such changes that undesirable consequences arise from favorable origins, and desirable consequences from unfavorable origins. What could have been more unpropitious to our prosperity than that, while He was engaged in reforming Russia, PETER and the country were threatened by attacks from without, by afflictions from within, and by dangers on all sides, and dire consequences were brewing? The war hampered domestic affairs, and domestic affairs hampered the war, which worked injury even before it had started. The Great Sovereign set out from His native land with a great embassy to see the states of Europe and to acquire knowledge of their advantages, so that on His return He might use them for the benefit of His subjects. Hardly had He crossed the frontiers of His dominion than He everywhere encountered great obstacles which had been set up secretly. However, I do not mention them now, since they are known to all the world. It seems to me that even inanimate objects sensed the oncoming danger to Russia's hopes. The waters of the Dvina sensed it and opened a path for their future Master amid the thick ice to save Him from the cunning snares that had been laid. Pouring forth, they proclaimed to the shores of the Baltic the dangers which He had overcome. Having escaped from danger, He hurried on his joyous path, finding pleasure for His eyes and heart and enriching His mind. But alas! He unwillingly cut short His glorious course. What conflict lie suffered within Himself! On the one hand was the pull of curiosity and of knowledge needful to the homeland; on the other hand there was the homeland itself, which had fallen on evil days and which, holding forth its hands to Him as to its only hope, exclaimed: "Return, return quickly; I am rent within by traitors! Thou art traveling in the interests of my happiness; I recognize this with gratitude, but do Thou first tame those who are raging. Thou hast parted from Thy household and with Thy dear ones to increase my glory; I acknowledge it eagerly, but do Thou allay the dangerous disorder. Thou hast left behind the crown and scepter given to Thee by God and hidest the rays of Thy Majesty behind a humble appearance for the sake of my enlightenment; with joyous hope do I desire it, but do Thou ward off the dark storm of turbulence from the domestic horizon." Affected by such feelings, He returned to calm the terrible tempest[46]. Such hindrances impeded our Hero in His glorious exploits! How many enemies surrounded Him on all sides! From abroad war was made by Sweden, Poland, the Crimea, Persia, many eastern nations, and the Ottoman Porte; at home there were the strel'tsy, the dissenters, the Cossacks, and the brigands. In His own household villainies, hatred, and acts of treachery against His most precious he were fomented by His own blood[47]. To describe all in detail would be difficult, and it would be painful to hear! Let us return to the joy of a joyous era. The Almighty helped PETER to overcome all grievous obstacles and to exalt Russia, fostering His piety, sagacity, magnanimity, courage, sense of justice, forbearance, and industriousness. His zeal and faith in God in all His enterprises are well known. His first joy was the Lord's house. He was not just a worshiper attending divine service, but Himself the chief officiant. He heightened the attention and devotion of the worshipers with His monarchic voice; and He would stand somewhere away from the sovereign's place, by the side of ordinary choristers, before God. We have many examples of His piety, but one shall now suffice. Going out to meet the body of the holy and brave Prince Alexander[48], He moved the whole city and moved the waters of the Neva with an act full of devotion. Wondrous spectacle! At the oars were the Bearers of various Orders, while in the stern the Monarch Himself was steering and, before all the people, lifting His anointed hands to the labor of simple men in the name of His faith. Made strong by His faith, He escaped the frequent assault of bloodthirsty traitors. On the day of the battle of Poltava the Lord shielded His head with power from on high and did not permit the death-dealing metal to touch it! As the Lord had once crumbled the wall of Jericho, He crumbled before PETER the wall of Narva --not while blows were being struck from fire-breathing engines, but during divine service.

---

[43] A distinct feature of Peter's financial policy was to build a budget on the country's internal resources without external loans.

[44] The Senate was established in 1711, and the Synod in 1721. The first nine Collegiums (Colleges, government departments) were established by Peter on the Swedish model in 1717 - the College for Foreign Affairs, the State Chamber (for taxation), the College of Justice, the College of Comptrollers, the War College, the Admiralty, the College of Commerce, the Treasury (for state expenditures), and the College of Mining and Manufactures.

[45] Established in 1722 and sorted all State servants in twelve ranks.

---

[46] Reference to the strel'tsi mutiny of 1698, see footnote 23.
[47] Peter's sisters, Sofia Alekseevna (1657-1704, Regent 1682-1689), Martha (1657-1707), Theodosia (1662-1713), his first wife Evdokia (1669-1731) and son Alexei Petrovich (1690-1718) were accused or suspected of planning against the tzar on different occasions and punished.
[48] St. Alexander Nevsky (1221-1263), Prince of Novgorod, Grand Prince of Kiev and Vladimir, canonized by the Russian Orthodox Church in 1547 for his leadership in defense against German and Swedish invaders. During a solemn transfer of his relics from Vladimir on August 30, 1724, Peter met them on the Neva river, personally transferred them to his galley and brought to St. Petersburg's church of newly established Nevsky Monastery.

Sanctified and protected by piety, He was endowed by God with peerless wisdom. What seriousness in counsel, unfeigned brevity of speech, precision of images, dignity of utterance, thirst for learning, diligent attention to prudent and useful discourses, and what unwavering intelligence [showed] in the eyes and entire countenance! Through these gifts of PETER, Russia took on a new appearance, foundation was laid for the arts and sciences, missions and alliances were instituted, the cunning designs of certain powers against our country were thwarted, while among Sovereigns some had their kingdoms and autocratic rights preserved, and to others the crown wrested from them by their foes was restored. Complementing the wisdom lavished on Him from on high and clearly manifested in everything which has been said above was His heroic courage; with the former He astounded the universe, with the latter He struck fear into His enemies. In His most tender childhood He showed fearlessness in military exercises. When all the observers of a new enterprise—the throwing of grenades onto a designated place—were exceedingly fearful of injury, the young Sovereign strove with all His might to watch from nearby, and was scarcely restrained by the teats of His Mother, the pleas of His Brother, and the supplications of dignitaries[49]. On His travels in foreign states in pursuit of learning, how many dangers He scorned for the sake of Russia's renewal! Sailing over the inconstant depths of the sea served him for entertainment. How many times the waves of the sea, raising their proud crests, were witnesses of unblenching daring as, cleft by the swift-running fleet, they struck the ships and combined with raging flame and metal roaring through the air into a single danger, but Failed to terrify Him! Who can without terror picture PETER flying over the fields of Poltava amidst His army, drawn up for battle, a hail of enemy bullets whistling around His head, His voice raised aloft through the tumult urging His regiments to fight bravely. Nor couldst thou, sultry Persia, halt our Hero's onset with thy swift rivers, miry swamps, high mountain precipices, poisonous springs, burning sands, or sudden raids of turbulent peoples, as thou couldst not stop His triumphal entry into cities filled with hidden weapons and guile[50].

For the sake of brevity I offer no more examples of His heroic spirit, O my Hearers, and I make no mention of the many battles and victories that occurred in His presence and under His leadership; but I do speak of His generosity, the generosity characteristic of great Heroes, which admits victories and moves the human heart more than bold deeds. In victories the courage of the soldiers, the help of allies, and the opportuneness of place and time have a share, and chance appropriates the greatest share, as it taking what belongs to it. But everything belongs to the conqueror's generosity alone. He wins the most glorious victory who can conquer himself. In [this victory] neither soldiers, nor allies, nor time, nor place, nor yet that chance itself which rules over human affairs have even the slightest portion. True, reason admires the victor, but it is the heart which loves the man of generous spirit. Such was our Great Defender. He would lay down His wrath together with His weapons, and not only were none of His enemies deprived of their lives merely because they had borne arms against Him, but they were even shown honor beyond compare. Tell me, you Swedish generals who were captured at Poltava, what did you think when, expecting to be bound, you were girt with your own swords which you had raised against us; when, expecting to be put into dungeons, you were put to sit at the Victor's table; when, expecting mockery, you were hailed as our teachers. What a generous Conqueror you had!

---

[49] According to legend, twelve-year-old Peter in 1684, being present at the Moscow Cannon Court, brought the wick to one of the firing cannons.

[50] Reference to Peter's Persian campaign and the capture of Derbent on August 23, 1722.

Justice is akin to generosity and is often linked with it. The first duty of rulers set over nations by God is to govern the realm in righteousness and truth, to and reward merit and punish crime. Although military matters and other great concerns and, in particular, His premature death much hindered the Great Sovereign in establishing immutable and clear laws for everything, yet how much labor He expended in this matter is testified beyond doubt by the many decrees, statutes, and regulations, the drafting of which deprived Him of many days of relaxation and many nights of sleep. God has willed that a Daughter like unto so great a Sire should complete these things and bring them to perfection during Her untroubled and blessed rule.

But though justice was not established to perfection in clear and systematic laws, yet was justice inscribed in His heart. Though not everything was written in books, it was carried out in fact. At the same time mercy was favored in the courtroom in those very instances when villainies which had hindered many of His deeds seemed to compel severity. Of many examples, one will serve. Having forgiven many distinguished persons their grievous crimes, He proclaimed His heartfelt joy by taking them to His table and by firing cannon. He is not made despondent by the execution of the strel'tsy. Imagine to yourselves and reflect on what zeal for justice, pity for His subjects, and His own danger were saying in His heart. Innocent blood has been shed in the houses and streets of Moscow, widows are weeping, orphans are sobbing, violated women and girls are moaning, my relatives were put to death in my house before my eyes, and a sharp weapon was held against my heart. I was preserved by God, I endured these things, eluded danger, and took my way outside the city. Now they have cut short my beneficial journey, openly taking up arms against the homeland. If I do not take revenge for all this, averting eventual doom by means of execution, I foresee town squares filled with corpses, plundered houses, wrecked churches, Moscow beset by flames on all sides, and my beloved country plunged into smoke and ashes. For all these disasters, tears, and bloodshed God will hold me answerable. Mindful of this ultimate judgment, He was compelled to resort to severity.

Nothing can serve me so well to demonstrate the kindness and gentleness of His heart as His incomparable graciousness toward His subjects. Superbly endowed as He was, elevated in His Majesty, and exalted by most glorious deeds, He did but the more increase and adorn these things by His incomparable graciousness. Often He moved amongst His subjects simply, countenancing neither the pomp that proclaims the monarch's presence nor servility. Often anyone afoot was free to meet Him, to follow Him, to walk along with Him, to start a conversation if so inclined. In former times many Sovereigns were carried on the shoulders and heads of their slaves; graciousness exalted Him above these very Sovereigns. At the very time of festivity and relaxation important business would be brought to Him; but the importance did not decrease gaiety, nor did simplicity lessen the importance. How He awaited, received, and greeted His loyal subjects! What gaiety there was at His table! He asked questions, listened, answered, discussed as with friends; and whatever time was saved at table by the small number of dishes was spent in gracious conversation. Amid so many cares of state He lived at ease as among friends. Into how many tiny huts of craftsmen did He bring His Majesty, and heartened with His presence His most lowly, but skilled and loyal, servants. How often He joined them in the exercise of their crafts and in various labors. For He attracted more by example than He compelled by force. And if there was anything which then seemed to be compulsion, it now stands revealed as a benefaction. His idea of relaxation was to change His labors. Not only day or morning but even the sun at its rising shone upon Him in many places as He was engaged in various labors[51]. The business of the governmental, administrative, and judicial offices instituted by Him was a carried on in His presence. The various crafts made speedy progress not only through His supervision but also through the assistance of His hands: public buildings, ships, harbors, fortresses ever beheld Him, having Him as a guide in their foundation, supporter in their labor, and rewarder on completion. What of His travels or, rather, swift-soaring flights? Hardly had the White Sea heard the voice of His command before it was already felt by the Baltic; scarcely had His ships' wake disappeared on the waters of the Sea of Azov before the thundering Caspian waves were making way for Hint. And you great rivers, the Southern Dvina and the Northern Dvina, Dnieper, Don, Volga, Bug, Vistula, Oder, Elbe, Danube, Seine, Thames, Rhine, and others, tell me, how often were you granted the honor of reflecting the image of PETER the Great in your waters? Will you tell me? I cannot count them! Now we can only contemplate with joyous amazement the roads along which He went, the tree under which He rested, the spring that quenched His thirst, the places where He labored with humble persons as a humble workman, where He wrote laws, where He made plans of boats, harbors and fortresses, and where at the same time He conversed with His subjects as a friend. In His care and labors for us He was in constant motion, like the stars of the sky in their course, like the ebb and flow of the tide.

In the midst of fire in the battlefield, amid weighty government deliberations, amid the diverse machinery of various crafts, amid a numerous multitude of peoples engaged in the building of towns, harbors, and canals, amid the roaring breakers of the White, Black, Baltic, and Caspian seas and of the very Ocean itself—wherever I turn in spirit, everywhere do I behold PETER the Great in sweat, in dust, in smoke, in flame; and I cannot convince myself that it is not many PETERS everywhere, but a single one, not a thousand years, but one short life. With whom shall

---

[51] According to (Stälin J. von. Originalanekdoten von Peter dem Grossen. Leipzig, 1785, p. 325) in the wintertime Peter woke up at 4 a.m. and immediately got down to business, at 6 a.m. he went to the Senate, to the Admiralty or other institutions.

I compare the Great Sovereign? In both ancient and modern times I behold Rulers termed great. And in truth they are great when compared with others; but compared with PETER they are little. One has conquered many states but has left his own country untended. Another has vanquished a foe who was already called great, but has spilled the blood of his citizens on all sides solely to gratify his own ambition and instead of a triumphal return has heard the weeping and lamentation of his country. Some were adorned with many virtues but, instead of lifting up their country, were unable to keep it from sinking. Some have been warriors on land but have feared the sea. Some have ruled the waves but feared to put in to shore. Some have loved learning but feared the drawn sword. Some have feared neither steel nor water nor fire, but have lacked understanding of man's estate and heritage. I shall quote no examples except that of Rome. But even Rome falls short. What was achieved in the two hundred and fifty years from the First Punic War to Augustus by Nepos, Scipio, Marcellus, Regulus, Metellus, Cato, and Sulla —as much was achieved by PETER in the short period of His life. Then to whom shall I liken our Hero? I have often pondered the nature of Him whose all-powerful hand rules sky, land, and sea. Let His breath blow and the waters shall pour forth; let Him touch the mountains and they shall be lifted up. But a limit has been set to human thoughts! They cannot grasp the Deity! He is usually pictured in human form. And so, if a man must be found who, in our conception, resembles God, I find none excepting PETER the Great.

For His great services to the country He has been called Father of the Country. But the title is too small for Him. Say, what name shall we give Him in return for begetting His Daughter, our most gracious Sovereign, who has ascended Her Father's throne in courage, vanquished proud enemies, pacified Europe, and lavished Her benefactions on Her subjects?

Hear us, O God, and reward us, O Lord! For PETER'S great labors, for the solicitude of CATHERINE, for the tears and sighs that the two Sisters, PETER'S two Daughters, poured forth when taking Their farewell[52], for all Their incomparable benefactions to Russia, reward us with length of days and with Posterity.

And Thou, Great Soul, shining in eternity and casting Heroes into obscurity with Thy brilliance, do Thou exult. Thy Daughter reigns; Thy Grandson is heir; a Great-grandson has been born in accordance with our desire; we have been exalted, strengthened, enlightened, and adorned by Thee; by Her we have been delivered, enheartened, defended, enriched, glorified. Accept as a sign of gratitude this unworthy offering. Thy merits are greater than all our efforts.

---

[52] Reference to the 1727 parting of then Grand Princess Elizaveta Petrovna with her sister, the Duchess of Holstein Anna Petrovna (1708-1728, mother of Peter III).

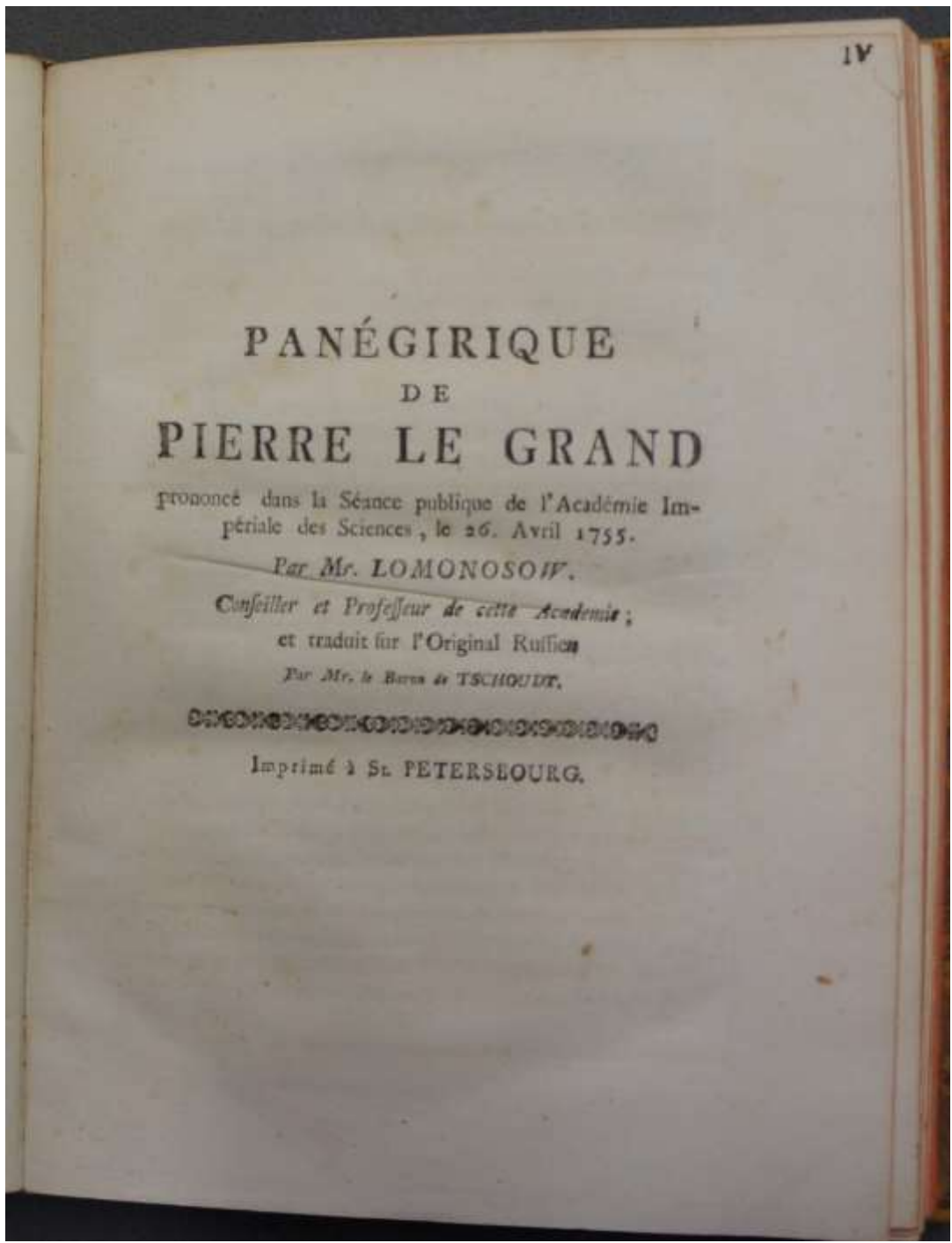

Fig.1: Front page of the "*Panegyric to the Sovereign Emperor Peter the Great*" (from the convolute *Lomonosow Opera Academica* in the collection of the Bologna Academy of Sciences library).

*Commentary* [by V.Shiltsev]:

This "Panegyric to the Sovereign Emperor Peter the Great" [1], written and read in 1755, continues the series of English translations of Mikhail Lomonosov's seminal works which were included by himself in the convolute *Lomonosow Opera Academica* intended for distribution among European Academies. It is the fourth article among the total of nine – see Fig.1. Translations of some of papers available at present can be found in Refs. [3, 4, 5] and two others – "Oration on the Use of Chemistry", "Oration on the Origin of Light" - in the Henry Leicester's book [6]. This English translation of Lomonosov's oration is a minor tune-up of generally excellent translation by Ronald Hingley [7], with only a dozen of notable textual changes. Lomonosov's 18[th] century Russian - language of the original publication [1]  - is not an easily to read and to understand by the modern day Russian. *Opera Academica* contains French translation by Baron Thoedore-Henri de Tschoudy (1727-1769), secretary of Lomonosov's patron Count Ivan I. Shuvalov (1727-1797). That translation of 1759 was sent by Shuvalov to Voltaire (who acknowledged getting it while working on the *Histoire De L'empire De Russie Sous Pierre Le Grand*), though according to Lomonosov, it was of poor quality and contrary to the author's objections mostly related to several textual "improvements" by the translator. In 1761, the "Panegyric…" got high praise from Johann Christoph Gottsched *(1700-1766)* in whose books and articles Lomonosov once studied poetry and rhetoric. Gottsched published a German translation in *Das Neueste aus der anmuthigen Gelehrsamkeit* (1761, pp. 200-207) with accompanying words: "Now our readers can judge on their own what a courageous power and a good taste this Russian poet possesses." In Russia, the "Panegyric…" had great success, additional prints were requested in distributed in 1755, second edition was published in 1757 and it served as a model for panegyrics to Peter in the second half of the 18th century. More on the life and works of the outstanding Russian polymath and one of the giants of the European Enlightenment can be found in books [8, 9] and recent articles [11-14].

The theme of Peter the Great [15] (1672-1725, see Fig.2) was central to Lomonosov's activities from the very beginning. The reviews on Peter deeds, accomplishments and ideals are scattered across different texts of Lomonosov, as well as references to Peter, appeals to him,, etc. All that apparently reflects the constant presence of the Petrine ideal, that Lomonosov never tired of admiring. Notably, both champions of Russian Enlightenment looked very much alike, and there was a remarkable and revealing 20[th] century legend that Lomonosov was the illegitimate son of Peter the Great (that is not true – see Ref.[11]). The occasion of the panegyric was the 14[th] anniversary of the 1741 coronation of Peter's daughter Elizabeth, and its purpose was clearly to remind the empress of her farther policies, most notably on his zeal on dissemination of technologies, crafts and arts and on tireless support of sciences in general and St. Petersburg Academy (which was the founder of, see, e.g., Refs.[16-18]), in particular – see Fig.3. The panegyric was important not only as a sympathetic retrospective review of Peter's reign, but also as one of foundational contributions to the myth of Peter the Great as a heroic man larger than life and the founder of modernized Russia.  Considering himself to be a direct continuation of Peter's works, Lomonosov was sincere in his hyperbolic praises of the reformist.

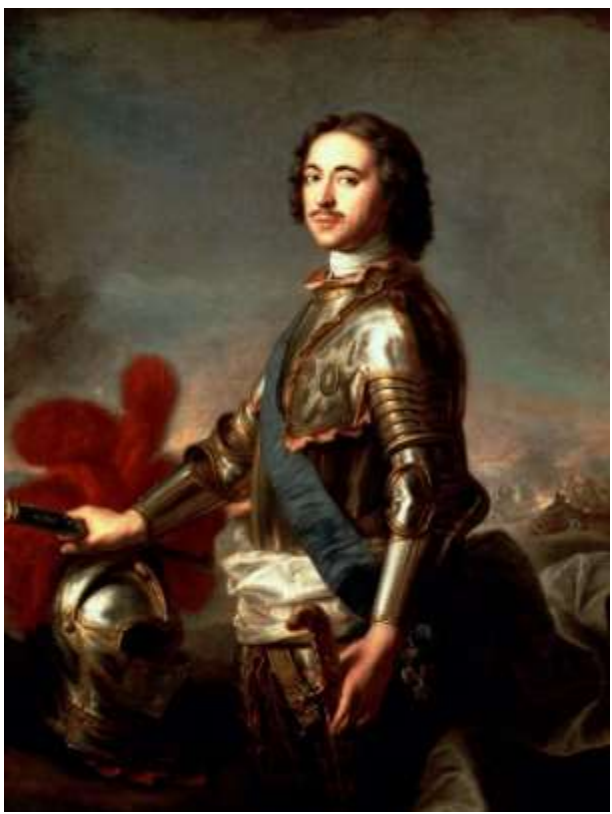

Fig.2: Portrait of Peter I by Jean-Marc Nattier (1685 – 1766). France, 1717. The State Hermitage Museum, St.Petersburg, Russia.

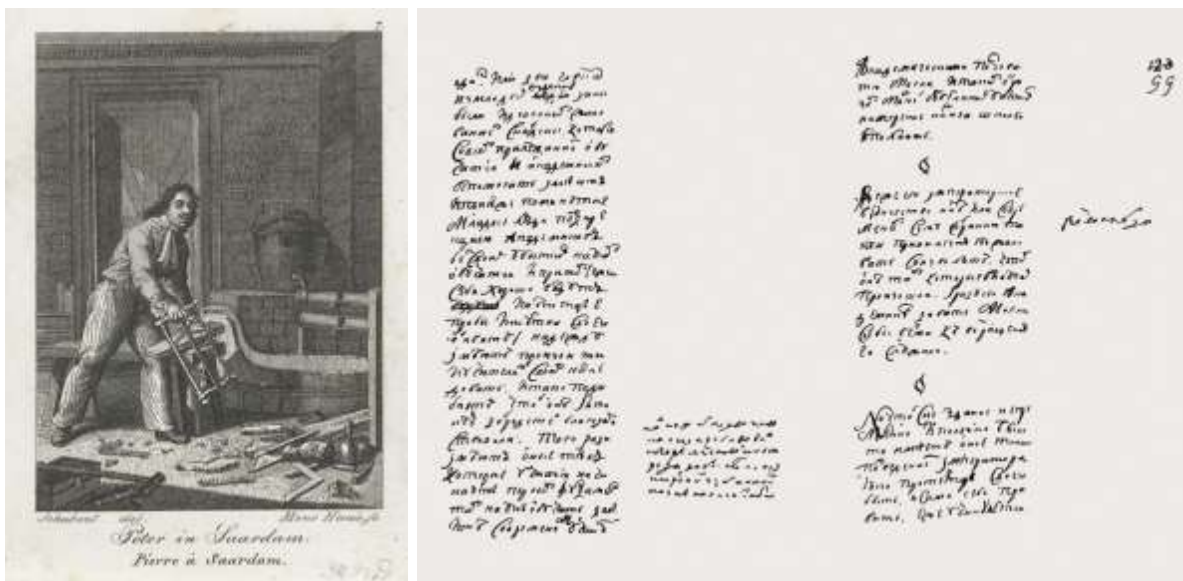

Fig.3: (left) Tzar Peter in Saardam, Holland in 1697 *(*Rijksmuseum, Amsterdam); (right) the 1724 *Ukazee* (decree) of the Governing Senate on establishment of the Academy of Sciences with Peter's handwritten corrections.

The panegyric is supposedly written in the period from 1754 (no later than November) to April 14, 1755. It was originally conceived by Lomonosov as "…laudable word on Peter the Great in a public Academic meeting in the next 1754, on which I am ready to put all my strength to". More than a year later, in 1754, Lomonosov finally got an official request from the Academy President Count Kirill Razumovsky (1728-1803) and at the end of November 1754 this work was in full swing. By early December, the first part of the panegyric was written and presented to the President of the Academy, who ordered to translate it into French and print 100 copies of the translation. The work then stalled due to delay of the Academy public assembly and complicated internal Academy politics related to the revision of the Academy Charter. Lomonosov initiated a campaign "to correct" the Academy. The situation at the Academy became so difficult for Lomonosov that he even requested to retire and to transfer him to some other State department. In February 1755 the President had forbid Lomonosov to attend the Academic assembly (internal conference). The ban was lifted at the end of March 1755 under pressure from senior patrons of Lomonosov, who was ordered "… to try as much as possible to prepare for the future assembly, that is, on April 26, the commendable oratory to Emperor Peter the Great, which he had promised to compose for a long time". Over the next two weeks, the "Panegyric" was finished, the Academic Chancellery issued an order to print 600 copies, distribute 150 and sell the remaining 450 copies at 40 kopecks per copy, the printing was finished by Academy's Print House in time.

For four days prior the Assembly, Lomonosov prepared and rehearsed his speech: memorized the text, tuned up the "intonations" proper for the expected audience of the highest ranks of the Empire. The oratorical success of the "Panegyric" was beyond doubt: the ceremonial bound copies prepared for distribution to "noble persons" were not enough and additional 50 copies were bound.

On April 28, 1755, the *St. Petersburg Vedomosti* (central newspaper) informed about the April 26th "…public meeting at the Academy of Sciences, and the College Adviser and Chemistry Professor, Mr. Lomonosov, gave a laudable oratory in Russian on the blessed and everlasting memory of the Emperor Peter the Great in presence of a large meeting of notable ministers, courtiers and other noblemen."

*I would like to thank Prof. Robert Crease of SUNY, my long-term collaborator and co-author of several scholar papers on Mikhail Lomonosov, for the encouragement to translate Lomonosov's major works to English. Special thanks to Profs. Luisa Cifarelli and Aldo Roda (Bologna) who provided me a unique opportunity to visit the Bologna Academy library and archives and access Lomonosov's convolute and correspondence.*

[14] Crease, R., Shiltsev, V. "Mikhail Lomonosov (1711-1765): Scientist in Politically Turbulent Times", *Il Nuovo Saggiatore*, v.33, No 5-6 (2017), pp.43-55.

[15] Massie, R. *Peter the Great: His life and world* , Random House (2012).

[16] Boss, V. *Newton and Russia: The Early Influence, 1698–1796* (Cambridge, MA: Harvard University Press, 1972)

[17] Gordin, M. ''The Importation of Being Earnest: The Early St. Petersburg Academy of Sciences,'' *Isis* 91, no. 1 (2000), pp 1–31.

[18] Lipski, A. ''The Foundation of the Russian Academy of Sciences,'' *Isis* 44, no. 4 (1953), pp. 349–54

# Mikhail Vasil'evich Lomonosov

## *Panegyric to Empress Elisabeth* [1]

### *August autocrat of Russia; presented by Mikhail Lomonosov on November 26, 1749. [Latin translation by the same author.]*

*[ PANEGYRICUS ELISABETAE AUGUSTAE RUSSIARUM IMPERATRICI PATRIO SERMONE DICTUS ORANTE MICHAELE LOMONOSOW. LATINE REDDITUS EODEM AUCTORE ]*

*[Слово похвальное Ея Величеству Государыне Императрице Елисавете Петровне, Самодержице Всероссийской, говоренное Ноября 26 дня 1749 года ]*

  If on this the glorious and solemn holiday, my Listeners, when all the people peacefully living under the blessed auspices of our most gracious Monarchess have the greatest tranquility and celebrate triumphantly her glorious ascent to All-Russian throne, it would be possible for us, delighted with joy, to fly to the great height from which we could watch the vastness of Her Empire and to hear from the ascending to the setting sun unceasingly extending exclamations and the air being filled with the name of Elizabeth – then what a beautiful, what a magnificent joyful view would have opened! Our spirit would have enjoyed so many various views of the celebrating ones and we could feel with our senses that the cities in stronger peace than if fenced by walls, in the villages blessed by fertility, in the seas free from military storms and noises, on the rivers abundantly flowing between merry beaches, in the fields decorated by contentment and safety, on the mountains raising their peaks in well-being, and on the hills girdled with joy, different inhabitants with different altarpieces, different ranks in different magnificence, different tribes in different languages extol to the one, they rejoice about the one, and they boast the only merciful their Autocrat. There, with reverence standing in front of the altar to the Lord, the deacon elevates the prayer voices with smoking oliban and his heart to God for Her covering and adorning his church in a deep tranquility; and in other places joyful sounds of peaceful weapons reach the clouds with solemn applauds of the Russian army, showing their zeal for their prosperous and generous Monarchess. There, having come together for a festive feast, the governors and citizens in adoring conversation recall the works of Peter, now being performed with the vigor by his Augustine Daughter; in other places after fruitful summer and at full breadbaskets the farmers dance and with their simple, but eager singing, glorify their Patroness. There, the sailors, resting in a safe haven, remember the storms with joy and celebrate this day with utmost joy; elsewhere, horse-riding along the vast fields of Asia, the steppe inhabitants shoot arrows from bows with their tricky art and show how they are ready to point them on the enemies of their Sovereign.

---

[1] Translated to English from the Russian version of Ref. [1] by V.Shiltsev; footnotes and commentary by V.Shiltsev

But, although the natural limits of human forces do not allow our joyful sight to achieve such elevations and to rejoice all that with our eyes, however, we ascend with the spirit, with vigorous wings of thoughts we fly up and we see general celebration everywhere with our mind's eyes, which, most of all, extend to the ancient royal city, glorified by the presence of Our Monarchess. Often our mental gaze, overlooking various views of celebrations, adorning Her blessed possessions on this day, turns to Her serene face and finds all kinds of signs on it alone. It shows true piety, delighting the church, on it is courageous cheerfulness, strengthening the army, on it is angelic justice, giving example to judges and hope to those being judged, foresighted wisdom on it that overlooks distant places and future times is clearly seen in her absence and equally reverentially honored in her presence. But who, with earnest zeal of eyesight, could see this more clearly then the society for the spread of sciences in Russia established by Peter the Great and renewed by her unspeakable generosity[2]? Neither mountains nor forests can screen Her divine eye that is inscribed in our souls. Her sweetest mouth, commanding to encourage us, and her eyes, humanely shining to us, and a generous hand[3], signing our well-being. The encouragement of the dawning sciences, not sparing her care, approval of their well-being by prescribing useful laws, protecting them with grace via accepting under own patronage, allowing them free access to herself, having entrusted them to the good-willed Presenter from her closest ones – that is such a great blessing, which will forever remain indelibly in our thoughts and hearts and for which we, at every opportunity and within our strength striving to strengthen the sciences and extolling our great benefactor, must give praise by deeds and words.

But when should we be most excited to express our gratitude if not as on this solemn day enlightened by Her ascension on the Father's throne, on which our special joy adjoins the general celebration? Now our indescribable joy cannot be confined in the narrow limits of the hearts, but also pours on our faces and tongues. Extreme powers of the mind and words strive to depict Her Monarchess virtues, the amusement of her subjects, the wonder of the world, the glory and the decoration of our times.

I undertake a great deal beyond the power of my mind, when I begin to bring thanksgiving and praise for the inexpressible blessing of the greatest Empress in the world at the noble meeting on behalf of our scholarly society. But, judging diligently, I find it rather easy and soundly, for where is the richest subject the eloquence can find, where the mind can spread more widely, where sincere jealousy can rush faster, as in the most glorious virtues of such a great Monarchess? When my tongue, blessed with Her encouragements, can act more conveniently, when my voice fortified by Her generosity can rise louder, as when preaching and praising Her incomparable merits? Neither by circumlocutional proliferation of thoughts, nor by ornate addition of ideas or a motley expression of utterances, nor by rhetoric soaring will my word be elevated, but all its strength and majesty will get it from the incomparable features of our Monarchess, all its beauty from Her

---

[2] Meaning the St.Petersburg Academy of Sciences, established by Peter the Great in 1724, which got a new Charter in 1747.
[3] According to the 1747 Charter, state support of the Academy doubled.

beautiful virtues and all its elevation from sincere aspirations to Her, because the gratefulness is conveyed to the devote Monarchess  - as testify erected and ornamented temples of the Lord, feastings, prayers and difficult homages;   the gratefulness is conveyed  to the courageous Monarchess – as testify Her most glorious victories over inner and external enemies[4];   the gratefulness is conveyed to the magnanimous Monarchess – as testify forgiven internal crimes and the prowess of external enemies and Her meek punishment of villains[5]; the gratefulness is conveyed to the wise Monarchess – as testify her visionary undertakings asserting the inner and outer quietness; the gratefulness is conveyed to the humane Monarchess — as testify Her maternal clemency and loving meekness to subjects; the gratefulness is conveyed to the merciful Monarchess – as testify the innumerable remissions from death penalties and that the sword given to her from God for the execution of the guilty, has not yet stained with blood[6]; the gratefulness is conveyed to the benevolent Monarchess – as testify rich support of faithful, abundant awards for services, help to true poor and compensations to unluckily broken. My mind now turns in a pleasant and magnificent paradise, being distracted from one blossoming virtue by the beauty of the other! All are glorious, all are adorable. From all that is clear, how noble is the root from which these grapes of virtues have grown and flourished. From all the virtues of our Monarchess, one can see how  great were Her ancestors, as they revived, empowered, restored, fortified and enlightened Russia so it is now  exalting her head above all earthly kingdoms, and the glorious deeds and merits of Her ancestors to the fatherland are as worth of the praise of Her Majesty as their blood which served for Her birth. To praise them, I would describe to you young Michael[7], accepting among moaning and tears of our great-grandfathers the crown of the Tsar together the heavy burden of the defeated Russia, renewing the ruined walls, rebuilding busted temples, gathering spread out citizens, refilling the lost state treasures, outrooting the God abjuring predators of the Russian throne and healing Moscow from the cruel defeat and the deep wounds;  I would now depict the wondrous and courageous Alexei[8] whose vigorous spirit encouraged Russia muscle flexing, who affirmed the well-being of subjects by salvatory laws, of regiments by the military science, of the church by the extermination of heresy and who extended his victorious sword to Sarmatia and returned to Russia by the righteous weapon  great principalities belonging to it since ancient times; I would present Peter by the name of the Great[9], by the wisdom of God, who enlightened Russia by the divine wisdom and frightened the universe by courage, with one hand keeping sword and hag-taper and another expanding to crafts, out-mastering other Monarchs in government and his servants in labors, filling the land with new regiments and covering the sea with a new fleet,

---

[4] Reference to the 1742 victory over Sweden and Rhein campaign of 1748.
[5] "internal crimes" refer to the affair of Heinrich Johann Friedrich Ostermann and Burkhard Christoph von Münnich (exiled in 1742); "external enemies" – Swedes: in August 1742, the Swedish army, surrounded by Russian troops in Helsingfors, surrendered and was allowed to return to its homeland, while the Finnish army, mobilized by the Swedes, was allowed to disperse to their homes.
[6] On the night of coup-d'etat 25 November 1741, to-become-Empress Elizabeth vowed to refrain from death sentences.
[7] Tzar Michael (Fedorovich) Romanov (1596-1645), grandfather of Peter the Great
[8] Tzar Alexei (Mikhailovich) Romanov (1629-1676), father of Peter the Great
[9] Tzar Petr (Alexeevich) Romanov (1672-1725), Peter the Great, Emperor since 1721

asserting the military art by His own example and raising the glory of the fatherland up to heaven; I would inscribe in your minds the Heroine, the beautiful Augustine Catherine[10], giving the wise advice to the wise Emperor with unshakable spirit among the barbarian raids, among thundering weapons, among the rustling cannon balls, who was then crowned by Him and accepted on Her shoulders all the state affairs of the hardworking Russian Hercules after His death; but my word hurries is to Our Monarchess virtues and accomplishments: on them my words will exhaust the strength, without getting into details, but presenting the most remarkable ones only. For the sake of that, I do neither depict in words the brilliant beauty of Her face, revealing a beautiful soul, nor the dignified age appropriate for a decent Monarchess, nor the majestic head, born to bear a crown, nor the mouth exuding generosities, nor her eyes reviving all those whom the human-loving Empress turns Her looks at. Everyone sees, everyone portrays in his mind that same way Peter the Great looked at the renewing Russia; same way he spoke strengthening his army and encouraging the labors of his subjects; same way he stretched out his hands, establishing crafts and science, commanding to arrange regiments for battle and fleet to leave for the sea; same way he lifted up his head entering the defeated cities and trampling upon the defeated enemy weapons; similarly he walked cheerly, inspecting his abuilding walls, ships under construction and restoration and piers and fortresses rising from the bottom in the seas; I will not present the appearance of our Monarchess, but instead will try to portray Her inner spiritual virtues which are faced by God affably such as decency liked by pious people, strong affirmation of states, the beauty of the Tsar's crowns, the praiseful solidity in the battle, the unbreakable union of human society. Great disorders, battles and homicides among the nations of one blood and one language happen due to separation of faith, and, on the contrary, the unity of faith firmly strengthen mutual union of harts, and more by examples rather than by a doctrine. Russia is prosperous because she confesses one faith in one language, and is ruled by one pious Autocrat, seeing a great example in Orthodoxy Her. One sees everywhere the growing number of churches glittering like the stars of heaven and glistened by Her radiance; one sees in fascination that that the Sovereign of many states to whom the land, sea and air serve for pleasure, strengthened by faith, keeps strict fasts, exhausts her body, and undertakes long journeys to sacred places on feet while which not only magnificent chariots and selected horses are available, but also hands and heads of the sons of Russia are ready to wear her. What a zeal to the Supreme is inflamed in our hearts and what an undoubted mercy we await from Him when we see our Autocrat before our eyes, kneeling and praying with us with extreme reverence! How courageous the Russian warriors facing the enemies, when they know that God, strengthening them in battle, God, loving their most pious Monarchess - goes to battle with them! So admired are the sacred places being resorted by Her piety presence! Russian Church being adorned with Her holy eagerness, like a bride on the day of triumphant marriage, shines with porphyry and gold and more with joy, beaming to the brightest altar of His throne and, showing Him her splendor and surrounded by glory, proclaims: "As Thy beloved Elisabeth beautifies me on Earth – please, adorn Her state and crown the with non-fading kindness of glory; as She lifts up my spiel in the heavenly kingdom – please exalt Her over all the rulers of

---

[10] Empress Catherine I (1684-1727), wife of Peter the Great, mother of Empress Elizaveta Petrovna

the Earth; as She visits me with diligence — please, award Her relentlessly with Thy grace; as She affirms my pillars in Russia — please, affirm Her health unshakably; as She helps me in defeating impiety – please, help Her in conquering arrogant and envious enemies and bless over Her army by Your strength". All subordinates concur with this sacred voice of Holy Church; therefore we believe that the Lord – as invincible pious champion of the glory – with His hand guides the courage of our August Monarchess in all the undertakings and affairs so it could be opposed neither by entrenched enemies inside Russia nor external attackers. The latter were defeated in one summer, and the former were thrown down in one night. To accept the guarder crown of the fathers and the Scepter with strong hands and embraced Russia with great authority would be a great challenge barely surmountable even by the man's heart and by a great Hero. But our Heroine, led by God, with a small number of faithful sons of the fatherland, despised all obstacles, triumphed without bloodshed and got Her heritage to our common joy. A wonderful and beautiful vision is portrayed in my mind when I imagine how the Maiden comes with a cross and armed soldiers follow Her. She is inflamed with her fatherly spirit and faith in God, while they fervently strive for Her; She fulfills the desire of all Russians, and they are ready to make it happen; She, approaching victory, does not want a bloody victory, they zealously stand for Her in front of the whole world. And what is then? The oppressors gone dead seeing the coming of Peter's Daughter and an insensible weapon fell from trembling hands and bowed down before their legitimate Sovereign! The Monarchy was enlightened by Her entrance and the throne was enlighten by the accession, and the merry Russians thought that even the walls of Petropol were moving enliven by their joy. The treacherous shores of Baltic got scared, and haughty enemies who were already beginning to reach our borders, got numb and, horrified, reversed their envious gaze backward and cared more of retreat rather than of a battle. Peter the Great, living in his courageous Daughter, got depicted in their frightened minds, and their fathers appeared to them, lying in their blood in the fields of Poltava, and many thousands of their people taken prisoners and sent to a steppe half a world apart; and they imagined their cities and villages burning, Russian ships walking on the Earth as on the sea, Russian galleys and sailors coming out of the waves of sea. However, though the enemies got defeated at the walls of the Vilmanstrand, but the battle was fierce; and the brave Russian hands felt resistance, and the victory was bought by a considerable shedding of blood. But when the father's scepter and the sword were accepted the courageous Elisabeth, then, the enemies, like swept by some rapid breathing, turned to precipitate retreat and even with their defenses, with their strong walls, with impassable abatises and fast streaming waters, and not only they did not dare to resist, but they barely dared to look back, seeing that neither the swampy swamps, nor the mossy lakes, nor the steep rapids could impede the righteous Elisabeth's wrath and power of Her lightning-fast soldiers. Finally, they were so repressed from everywhere, so surrounded by sea and land from the Russian forces that if not a generous Victoria, none of them would have been saved, and there would be no one to announce about their death in their Fatherland, except via the sonorous glory of Her Majesty. This victory was even more so fascinating, as it seemed that Mars, imitating our gentle Empress, disliked to shed human blood, and all of Europe reasoned that Russia did not have a war with the enemy, but had just punished the deceitful for fury. This quality of our Grand Governess is property of the universe to deal with enemies very generously. Such was shown not only in the complete defeat of external enemies, but even greater example towards internal enemies during Her glorious entrance on the Father's throne. Embarrassed Russia resorted to Her and spoke with the voice of her chosen sons: "Take me into Your motherly arms, accept Your hereditary power and despise all the obstacles with your inborn cheerful Fatherly spirit. Trust in God - he will be the leader of your righteous enterprise. Trust in Yourself - You are the only true heir, You are the Daughter of my Enlightener. Trust in me - I will move all my strength to Your protection and through the heads and corpses of Your enemies I will open You the way to the throne ". But the magnanimous Governess instead preferred to stay off Her hereditary crown for a time, than to gain it by shedding of blood, and fearing more of the distress to the Fatherland than wanting the Majesty for herself, she finally leaned toward accepting the rule to preserve the state. Having risen to the heights of the authority, how does She revenge those who separated Her from Her legitimate inheritance, those who distressed by fierce pride and unscrupulous oppression? Those who deserve fierce death according to the law of God, according to state laws and at the request of the Russian people, are condemned by distancing from Her enlightened face, those unworthy of life are just imprisoned and the heroic deeds of Her great accession to the throne are so abundantly endowed with Her extreme generosities, which she is so richly gifted with that they do not fit the vastness of Russia, but also spills over to external nations. Sweden is defeated by Her weapon, but more is defeated by Her generosity; it fears Her invincible forces, but more impressed by the great and noble spirit, as the invincible Governess having gained such great advantages does conclude the piece with the defeated, like, more justly to say, forgiving the guilty ones. She who got all the power of the defeated enemy in Her hands and can fulfill all Her will over him, but gives everything back to the defeated and resurrects the exhausted rebel - does not She forgive more than reconcile? But the virtue of our Monarchess extends further, as the Russian heroine shows a greater example of generosity, because not only She lets the enemies away with their country, peace, quietness and returns their lands, but also spreads Her weapon to their defense, turns back the other country threatening them with a war and asserts the inheritance of their throne together with liberty[11]. Having these considerations in mind and looking at the flourishing condition of the Russian state, the abundance of our vast Fatherland and the moderation, with which our worshiped Empress rules Her people, is it possible for you, our neighbors, to conceive, that Her noble heart bows to the appropriation of foreign lands? How can one having an expanse of fruitful fields wish impassable marshes? Can the one stretching Her scepter over the flowing in Her obedience abundant rivers transcending great Nile over flattering mosses? Will the one prevailing over the honey and milk emanating land look with desire over the barren stones? That the brave Russian armies are prepared to armor, that the fleet is ready to cover the Baltic waters, that all military preparations are in time – all that foretells not the war coming

---

[11] After the Russian-Sweden peace of 1743, Denmark began to threaten Sweden with war. In this regard, at the request of the Swedish government, the Russian fleet was moved to the shores of Sweden to protect them, and two Russian regiments were transferred to Stockholm. At the same time, the candidate nominated by the Russian court, the Holstein prince Adolf-Friedrich, was recognized as the heir to the Swedish throne.

from Russia, but the wisdom of our far-sighted Heroine. A skilled navigator is active not only in terrible excitements and storms, but also during the meek quiet times – he strengthens his instruments, sets the sails, watches the stars, notes changes in the air, observes at the rising clouds, calculates the distance from the coast, measures the depth of the sea and reveals the hidden underwater stones. In a similar way, the wisest Elizabeth, while rejoices looking at her subjects enjoying the beloved peace bestowed by Her, cares about their future safety: protects them and those uncapable of war with arms outstretched over the Earth and over the sea and with a sharp eye illuminates the thoughts of those in battles; opens secret enemy cunnings coated by quiet jets of adulation; understands the past, considers the present and foresees the future. All that for the sake of if there are some of those envious of our well-being, who would dare to anger the peace-loving heart of our Monarchess with furious or insidious bitterness, then he will get in full about Her wise preparedness and, although he could be covered and separated from us by vast seas, great rivers or enclosed by mountains, he will experience his punishment and will think that the sea had dried up, the flows of the rivers stopped and the mountains, dropping, turned into flat fields; he will feel that it is not the Russian navy, but the whole of Russia arrived to his coasts. Rest in joy, beloved Fatherland, and enjoy serene century under the shelter of your wise Governess! So safe is your well-being! So incomparable to others is your bliss! Others with tears look at the smoking ruins of their cities torn apart by enemies, while you gaze with joy over the newly erected magnificent buildings ascending to the clouds. Others are embraced with fear, trembling day and night, seeing their citizens running after each other with bare swords and shedding blood of the uni-genous, while you are adorned by consonant allegiance of unanimous sons of the common Mother. Others suffer misery and hunger from the suppression of the merchants, from the destruction of the arts, from the trampling of agriculture and undergo, while here the roads welcome merchants, the docks are open, the markets filled with riches, the sciences and the arts increase, and your granaries are in abundance. Others, though they freed themselves of military thunder and fears, see the deplorable traces of their enemies, the sternly view of which are still clearly portrayed in their thoughts, while you rest in an uninterrupted calmness and disturbed by fears only by sleepy ghosts. This your dearest and most holy calmness is solely from wise care of your visionary Governess. Her providence and care provide you with abundance, cheer you with general concordance, enrich you with merchants and serene farming, adorn with your beloved peace and fill the Universe with your loud glory. This our perfect pleasure, general amusement, abundant enrichment, pleasant decoration, are multiplied by our incomparable Monarch's divine love of mankind, when - elevated to the height of power and majesty, which human power cannot exceed – comes to Her subjects with condescension beyond the reach of mortals. What could be more pleasant to the human heart and what could be more extraordinary in the world, as when the Empress, Master of the greatest part of the world, revered from all tribes and rulers of the earth, with gentle eyes, small talks and merciful acceptance allows Her servants to see Her? Yet we enjoy that kind of pleasant sight for all days. Our humanity-loving Empress differs from the great multitude of her surrounding subjects not by an arrogant look, not by a destructive voice, not by a terrible command, but by beautiful Majesty, quiet authority, noble forbearance and some divine power, infusing unspeakable joy into our hearts. The gates of Her blessed home are turned not by horror and awe, but gentle meekness, the mercy attracting the hearts of all and reliable guard of Majesty - the faithful love of the subjects. Those who enter it do not look around incessantly, horrified by the walls themselves, on the trembling feet, but they lightly walk into the sacred halls of Her house following their joy anticipation. There is no need to experience their innermost thoughts: the beauty of general pleasure is in their eyes, and the signed of their joyful hearts are written on their outstretched foreheads. What a pleasant feeling pours into hearts gazing at such indulgent Majesty! And what coolness flows into the blood of numb guilty ones, when they think of the mercy of the Empress which the human word can barely be enough to describe! Nothing can be as commendable as meekness, no other virtue of is kinder, nothing in human nature comes closer to the divine properties, like the forgiveness of the guilty and freedom from proper punishment. But where is the glory at the court bragged, where justice and mercy are more conspicuous, where accusation and forgiveness are more closely mated, where condemnation and freedom are mutually alike to each other, as before the highest Her Majesty the throne? Let others, depriving life, staining the sword with their own blood, diminishing the number of subjects, plunging torn human members before the people, trying to intimidate evil and vices to exterminate the evil, but our Mother of Russia corrects human morals with joyful examples. Others want to eradicate malice through strict and often inhuman punishment, but she instills virtue with generous reward. If someone, having a great garden, cares only about destruction of thorns, having forgotten fruitful trees and beautiful flowers to solder with the necessary moisture, he will soon see his trees dry and barren and flowers withered from the heat. On the other hand, he who irrigates the trees of fruitful and prosperous grass at a decent time, despising the tares and passing them by, he will enjoy the abundance of fruitful trees and will rejoice the beauty of flowers, which - having intensified - will dry up the obesity and the juices of useless and harmful vegetation, those growth and the root will stop. In a similar way, although it is useful to observe strictly the law against the guilty, but without rewarding virtue it is futile and more to depressing for the good ones than to correction of the evil ones; on contrary, rewarding virtue and praise for service alongside modest punishment of taints are enough to correct human mores, because the evil ones the wicked will feel despised and trampled and seeing an exalted virtue enjoying their righteous bribery, they get mortified by envy, or, turning to a zealous imitation of the same, they will try to commit themselves to get worthy. With such prudent mercy, the generous Empress tries to breed virtue and eradicate vices in widespread Russia! Punishes masterfully, monarchically provides, corrects without severity, rewards abundantly, resurrects with the deliverance of those who have transgressed, encourages the deserved ones by benevolences. My voice gets rather faint, my tongue becomes dull and my word scarce, to count the details of Her good deeds, however, this meeting established by Her dearest Parents is so pleased with the Empress's generous charity and so endowed that, with extreme gratitude, and zealousness, can find invent convenient ways for testimony of our pleasure and slavish sincerity. This blessing is all the greater, the more glorious and more worthy of the Peter's Daughter , as it is not only for us alone, not only for the youth studying here, but for every rank and title, to the entire Russian state, to the entire human race, for it is not only we, content with Her Majesty's bounties, some in the revelation of natural mysteries and in the study of the wondrous deeds of the all-wise Creator in tranquility we enjoy, and others, teaching instruction to students, with joy we feel the fruits of our labors; it is not only students who are nourished by Her abundant hand and doing studies free of worries of their daily needs, but also general well-being is offered. There is not a single place in Russia, enlightened by Peter, where science could not bear fruit; there is not a single person who could not expect benefit from them. What is holier and what can be more salvific, than, while learning in the deeds of the Lord, gaze mentally at the high glory of His throne and preach his majesty, wisdom and power? To this, Astronomy opens its spacious building: this entire visible world and wonderful deeds, Physics shows its diverse cunning, giving abundant and rich matter to the cognition and glorification of the Creator by creatures. In the sea, what is more necessary for travelers to different states, how to know the position of places, the course of rivers, the distance of hailstones, the size, abundance and proximity of different lands, the customs, customs and governments of different peoples? This is clearly shown by Geography, which subjects the entire universe to the vastness of a single view. What motivates military hearts to act courageously against enemies and to bravely defend the fatherland, as by glorious examples of great Heroes? This is recalled by History and Poetry, which, vividly describing the past deeds presenting them as the present ones: both expel the triumphant deeds of the great Sovereigns from the gloomy jaws of caustic antiquity. What does tame human passions and validate natural and civil laws? What can be better imagined as such a means to govern reason, show the direct way to action, tames human passions and affirms natural and civil laws? This is done by Philosophy. What is more dear to a person than his life and what is more amiable than health? Both are saved and continued by Medicine. What is more necessary in human society is the use of various machines and knowledge of the inner construction of things? These are revealed by Chemistry and arranged by Mechanics. All these are controlled by precise and careful Mathematics. All these used to serve to increase the bliss of man according to their usefulness, although differently. But all these will multiply in Russia, prosper and bear abundant fruits in due time because of the special beneficence of our generous Empress. A beautiful tree of wisdom will grow here, planted by Peter, protected by mercy and filled with generosity of a Daughter worthy of only a Parent, and will grow and its branches will spread throughout the entire universe. Opened with Her Majesty's rich hand, a wide door to the sciences into vast Russia, where they, in all contentment and in complete safety, will spread out, and acquire a new increment, new decoration, new enlightenment and new glory in new splendor at an unexpected height, at the very top of its the perfections of those set themselves to the whole world will show and with their full radiance will scatter the remaining night of barbarism from the most remote and still barely known places; for where it could be more convenient to accomplish stargazing and earth-measuring science, as in Her Majesty's vast country, over which the sun makes a whole half of its motion and in which every luminary, rising and setting, can be seen at the same time. Where it could be more possible to explore diverse types of natural things and phenomena, as in the fields, decorating great space with a variety of colors, on the tops and in the depths of the mountains, above the rising clouds and poured with various treasures, in the rivers flowing from the sultry India to the eternal ices, and on many vast seas, full of God's wondrous miracles, bowing their waves under Elizabeth's power? Where else can the Muses find the safest dwelling, as in vast and serene Russia, soothed by the insight of our Monarchine and protected by Her invincible power? Oh, what a great blessing will spread throughout the whole world from this generosity of our Monarchine! Oh, how desired is your well-being, Russian youths, who are fed by the grace of the generous Empress and exercise yourself in joyful labors! Imagine your future state, to which you have been chosen, listen with reverence that the Most August Empress, contenting you with her treasury, motherly commands: "Study diligently. I wish to see the Russian Academy, consisting of the sons of Russia; hasten to achieve excellence in the sciences: this is the benefit and glory of the fatherland, this is the intention of My Parents, this is what My will requires. The deeds of My ancestors have not yet been described and the great glory of Peter has not been fully praised yet. Prostrate in the enrichment of the mind and in the adornment of the Russian word. In My vast state, invaluable treasures, which nature abundantly pronounces, lie hidden and await only skillful hands. Exercise the utmost diligence towards the knowledge of natural objects and strive earnestly to earn My mercy. " Hearing this generous command of Her Majesty, strive, stay awake, keep up your course. And you, for whom the entrance to the sciences is freely opened, use this generosity in favor of your sons and the intentions of Peter, the care of Catherine, and do not abandon Elizabeth's generosity in vain. A dwelling place for the sciences was erected amid this reigning city not for vain, but so that the administrators of civil affairs from the places of justice, practicing military affairs from the walls of the Peterburg, standing in front of the Monarch's face from the blessedness of Her house, builders and managers of the Russian fleet from the heights of the ships and merchants from ships – all they should look at this building 47 in the midst of their exercises, think about the sciences and bow to them in love. 48 It is true that this beautiful home of muse, to our unspeakable extreme sorrow, sadness and contrition has lost its pleasant appearance to deplorable disgrace due to an unexpected accident from a formidable fire, on which we can hardly gaze without groaning and tears[12], but in this sorrow of ours we have one consolation, we take hope in the generosity of our all-merciful Empress, knowing that there is no such misfortune, there is no such misery that would be exceeded by Her generosity and would not be turned away by Her generous hand. So great 50 is the generosity of our incomparable Monarchess. The throne of All Russia is so adorned with virtue!  Such Monarchs are sent by God to earth when he has mercy on mortals; only the pious, when they deign to hear their prayers and receive their offerings; such the courageous and magnanimous, when he wants to throw down their enemies and put them to shame; such the wise, when wants to multiply their blessedness; such the philanthropic, such the merciful and generous, when wants to comfort them, expand and  gift. Flaunt these great gifts of the highest, Most Merciful Empress, and delight in Your divine blessings. Wherever Your glorious eye turns, everywhere the joyful faces of Your subjects, supported by Your generosity and living only by Your mercy, richly rewarded by You and exalted by You. The whole northern country, although at all times, but especially on this blessed holiday, after a fruitful summer and at the end of the blessed autumn, abundant with fruits from the earth, with riches from the sea, everywhere abounding by Your happiness, exalts Your glorious ascension to the fatherly throne with numerous triumphant voices, and these exclamations, which then came from sudden joy and from true love, are now repeated many times. Our indescribable pleasure and extreme gratitude cannot be depicted by any eloquence, however, we strive to testify to the best of our strength our sincere jealousy and slavish loyalty to our Majesty, 51 knowing that God and God's authority on earth have not so much hear for intricate rhetorical constitutions but instead look mostly at pure zeal.

---

[12] Reference to big fire 5 December 1747 after which the main building of the Academy, with Kunskamera and Library, was heavily damaged and got fully restored only two decades later.

*∗∗∗*

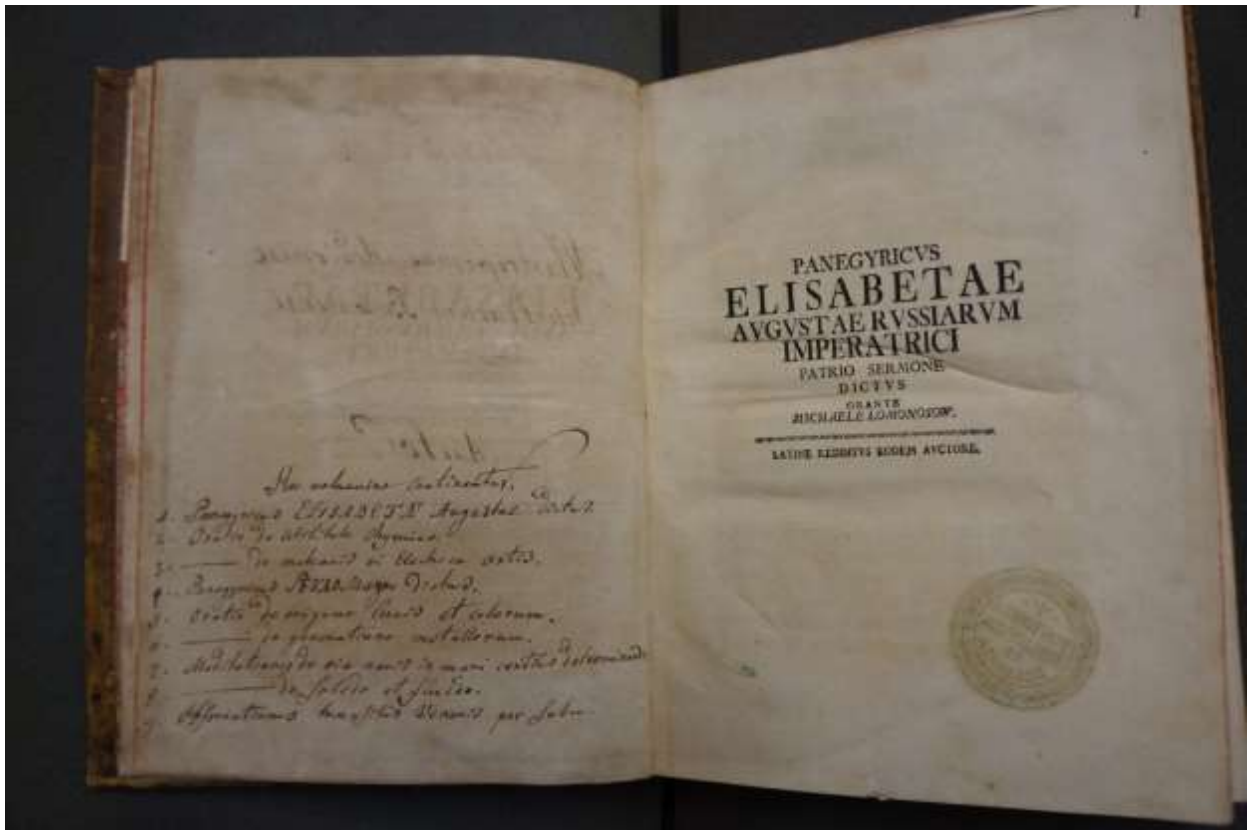

Fig.1: Front page of the "*Panegyric to Empress Elizabeth*" from the convolute *Lomonosow Opera Academica* with Lomonosov's handwritten table of content of the convolute (from the collection of the Bologna Academy of Sciences library).

*Commentary* [by V.Shiltsev]:

This "Panegyric to the Sovereign Empress Elizabeth" [1], written and read in 1749, continues the series of English translations of Mikhail Lomonosov's seminal printed works which were included by himself in the convolute *Lomonosow Opera Academica* intended for distribution among European Academies. It is the opening article among the total of nine – see Fig.1. Translations of some of papers available at present can be found in Refs. [3, 4, 5] and two others – "Oration on the Use of Chemistry", "Oration on the Origin of Light" - in the Henry Leicester's book [6]. More on the life and works of the outstanding Russian polymath and one of the giants of the European Enlightenment can be found in books [7, 8] and recent articles [9-14].

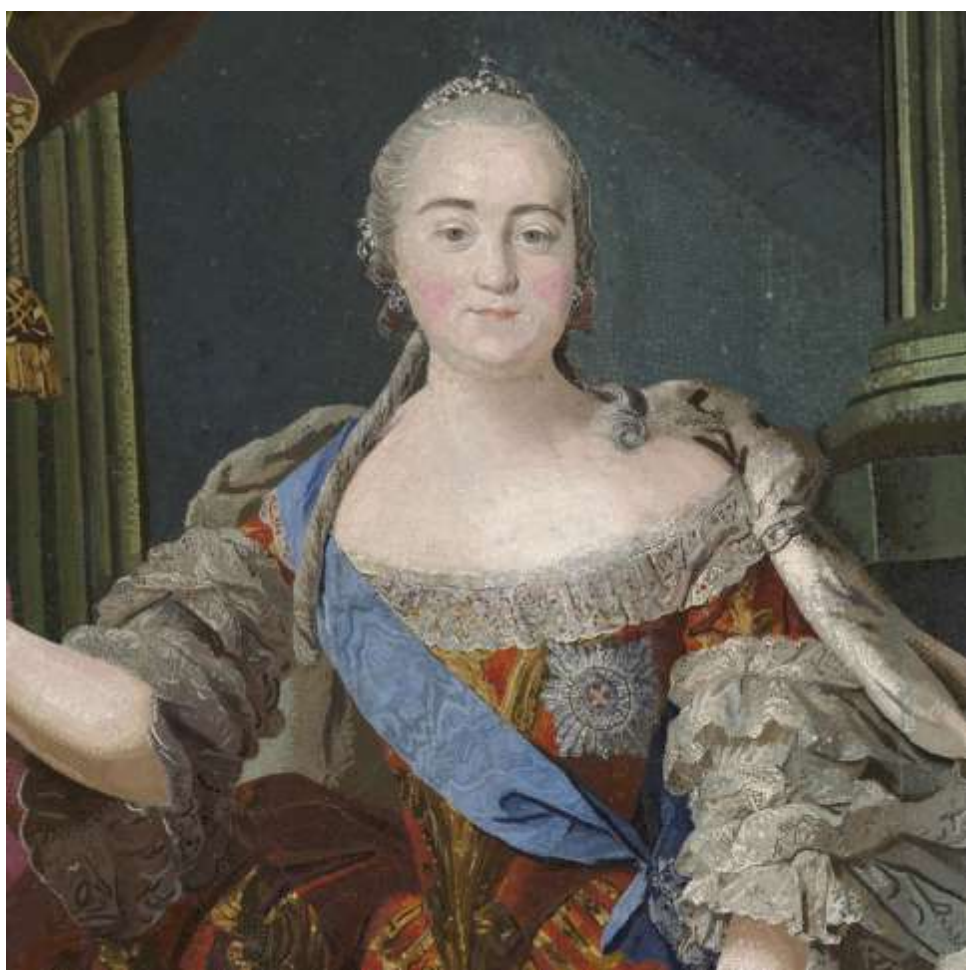

Fig.2: Mosaic portrait of Elizabeth I (Empress Elizaveta Petrovna, 1709-1762) by M.V.Lomonosov mosaics workshop. St.Petersburg, 1758-1760. The State Russkiy Museum, St.Petersburg, Russia.

This "Panegyric…" was unique in many aspects. It is the first major piece of oratory art by Lomonosov and it followed classical rules of the composition of panegyrics which Lomonosov himself developed following traditions of antiquity. Moreover, it was the first of a kind in the Russian literary tradition as well and set it up for many years. The panegyric was read in the public academic assembly on November 26, 1749, in the presence of the Russian Imperial court and Empress Elizabeth herself – see Fig.2. The assembly lasted three hours (11am-2pm) and had two parts – besides Lomonosov's oratory in Russian, Prof. G.Richman gave a talk "On the laws of evaporation" and Prof. C.Kratsenshtein responded to it at the behalf of the Academy (both – presumably, in Latin). According to the recently approved Chapter of the Academy (July 24, 1747), the Academic public assemblies were supposed to take place three time a year. So, the 1749 assembly was the first successful one after 5(!) of them got missed. Even that one was delayed (originally scheduled for the Elizabeth's nameday of Spetember 6) due to a scandal around originally scheduled scientific communication "On origins of the name and country of Russia" by Prof.G.Mueller. The latter supported so called "normanistic theory" which was found by many, including Lomonosov, factually incorrect and politically humiliating (after all recent Russian victories over Swedes). As the result, Mueller's talk was substituted with Richman's and Kratsenshtein's at the last moment. Fierce discussions between "normanists" (Mueller [15] and followers) and "anti-normanists" (Lomonosov [16] and followers) continued through centuries and are still active today [17].

This "Panegyric…" got high praise in Russia and abroad. Besides the 1749 publication, it was reprinted in 1751, 1755 and 1758. Copies of Latin translation (made by Lomonosov and published in parallel with the Russian version) were quickly sent abroad via diplomatic channels of Chamberlain Count Boris Yusupov (1695-1759). In his January 3, 1750 letter to Academic Chancellor J.-D. Schumacher Leonhard Euler (1707-1783) wrote from Berlin: "..i was very pleased to learn about great success of the latest public assembly of the Imperial Academy : all the oratories delivered on that occasions will deserve praise of the entire scientific world. That's especially true to the "Panegyric" by Mr. Lomonosov which seems to me to be exemplary in that kind ("*un chef d'oeuvre dans son genre*")." (AAN, f. 1, op. 3, no. 39, fol. 148).

*I would like to thank Prof. Robert Crease of SUNY, my long-term collaborator and co-author of several scholar papers on Mikhail Lomonosov, for the encouragement to translate Lomonosov's major works to English. Special thanks to Profs. Luisa Cifarelli and Aldo Roda (Bologna) who provided me a unique opportunity to visit the Bologna Academy library and archives and access Lomonosov's convolute and correspondence.*